\documentclass[aps,prper,reprint,amsmath,amssymb,floatfix]{revtex4-2}
\usepackage[T1]{fontenc}
\usepackage{graphicx}
\usepackage{booktabs}
\usepackage{xcolor}
\usepackage[hidelinks]{hyperref}

\newcommand{\exampleimage}[2]{%
  \IfFileExists{#1}{%
    \includegraphics[width=\linewidth,height=1.55in,keepaspectratio]{#1}%
  }{%
    \fbox{\parbox[c][1.45in][c]{\dimexpr\linewidth-2\fboxsep-2\fboxrule\relax}{%
      \centering\small #2\par\medskip
      \footnotesize\texttt{\detokenize{#1}}}}%
  }%
}

\newcommand{\plotfile}[2]{%
  \IfFileExists{#2}{\includegraphics[width=#1]{#2}}{%
    \fbox{\parbox[c][1.6in][c]{\dimexpr#1-2\fboxsep-2\fboxrule\relax}{%
      \centering\small Plot placeholder\par\medskip
      \footnotesize\texttt{\detokenize{#2}}}}%
  }%
}

\begin{document}

\title{What a difference two years make: AI-assisted scoring of a handwritten physics exam}

\author{Gerd Kortemeyer}
\email{kgerd@ethz.ch}
\altaffiliation[Also at ]{Michigan State University, East Lansing, MI 48823, USA}
\affiliation{Rectorate and AI Center, ETH Zurich, 8092 Zurich, Switzerland}

\date{\today}

\begin{abstract}
Handwritten physics exams combine equations, numerical calculations, diagrams, and lines of reasoning. Scoring this work, which may include incomplete, unexpected, duplicated, annotated, and scattered  derivations and drawings, pushes multimodal, reasoning-capable large language models to their practical limits. Those limits are shifting rapidly. We compare successive AI scoring systems on the same student solutions to a thermodynamics exam, drawing on studies published in 2024 and 2025 and a new analysis in 2026. At each stage, we sought the best practical performance from the state-of-the-art tools available to us, with particular attention to multimodal input and the amount of relevant context the model could consider at once. The comparison follows complete workflows, including transcription, rubrics, and confidence filtering. The results show substantially improved agreement with human scoring, especially for graphical work. Some remaining differences may reflect familiar judgments about the
interpretation of incomplete work and the allocation of partial
credit. Such differences emphasize the instructor's role in defining
what deserves credit and resolving disputed awards.
\end{abstract}

\maketitle

\section{Introduction}
\subsection{AI scoring of open-ended exams}
In a physics exam, a final numerical answer or correctly chosen option on a multiple-choice question is only part of the evidence of learning~\cite{lin1982learning}.
Credit should reflect how students represent a problem, identify relevant
concepts and assumptions, plan a solution, and evaluate their work,
even when they do not reach a correct final answer~\cite{Burkholder2020}. Research on the assessment of
written problem solutions has consequently emphasized both the structure of
the solution and the connection between physical and mathematical
reasoning~\cite{Docktor2016,Hull2013}. Handwritten work preserves much of this
evidence, but evaluating it traditionally requires substantial human attention.

Large language models offer a possible way to support this work without
restricting an exam to answers that can be checked by exact numerical or
algebraic matching. An early feasibility study combined handwriting
recognition with a language model to score physics derivations~\cite{Kortemeyer2023Grading}.
Subsequent studies investigated authentic handwritten thermodynamics exams
and ways to identify scores that should be referred to a human
grader~\cite{Kortemeyer2024,Kortemeyer2025}. Related work on written physics
explanations has shown that explicit rubric wording and repeated scoring can
improve agreement with human graders, and that variation across runs can help
locate problematic decisions~\cite{ChenWan2025}. A separate study found that the open-source models tested did
not meet the authors' accuracy requirement when extracting
and standardizing equations from typed physics
responses~\cite{mcginness2025can}.

Although a model's scoring and problem-solving performance
can be related~\cite{mok2025using}, the practical question
of using a particular model to score a physics problem
differs from asking whether it can solve that problem. A scoring system must determine what the student actually
wrote, distinguish intermediate work from an abandoned attempt, recognize
equivalent methods, and apply the intended partial-credit policy. A correct
solution generated by the model is of little use if it is mistakenly attributed
to the student. Conversely, a physically meaningful step should not lose
credit because it is written in an unexpected form.

\subsection{Human scoring of open-ended exams}
Scoring open-ended physics work requires judgments about what a written
solution demonstrates and how much credit that evidence warrants.
Instructors differ in how much reasoning they require students to show
and in their willingness to give the benefit of the doubt when reasoning
is incomplete or unstated~\cite{Henderson2004Grading}.
Independent graders can also assign systematically different amounts of
partial credit while producing strongly correlated
scores~\cite{SlepkovShiell2014}. 

Even in ``hard sciences'' like physics, scoring partially correct solutions requires judgment calls and is thus, to some level, subjective and disputable~\cite{speirs2026thematic}.
A rubric makes the criteria for credit explicit, but its construction
and application still involve choices. Work on the open-ended Colorado
Upper-Division Electrostatics diagnostic, which includes field sketches
and graphs, identifies both judgment in point allocation and the difficulty
of anticipating every student response~\cite{ChasteenEtAl2012}.
Response-category grading likewise makes the treatment of conceptual and
algebraic errors, and the credit attached to them, depend on the
instructor's learning goals~\cite{Paul2014ResponseCategory}.
These observations do not imply that disagreement is unavoidable:
rubric refinement, shared examples, and discussion among raters can
improve consistency~\cite{ChasteenEtAl2012,Docktor2016}, but in the end, the human factor is essential.

\subsection{Tracking an exam over the years}
How much scoring work can reasonably and responsibly be delegated to algorithms depends on their capabilities --- which in the case of large language models have rapidly evolved over what seems like an almost incredibly short time. We revisit one exam to ask how much the resulting scoring performance changed
over two years. At each stage, we used what we regarded as state-of-the-art
models and workflows available to us and sought to make the fullest practical
use of their multimodal and context capabilities. We did not intentionally
hold an older system to a minimal prompt or a restricted input when a more
capable workable configuration was available. Nevertheless, the study is not
an exhaustive ranking of all models (in our case, for legal and data security  reasons, we were limited to OpenAI models via the Azure AI Foundry), nor does it isolate the effect of the
model from improvements in the surrounding workflow.

\subsection{Research question}
We address three questions:
\begin{enumerate}
\item How did agreement with teaching-assistant (TA)
scores change for analytical, graphical, and numerical work?
\item How did the
relationship between AI and TA scores change for the complete exam?
\item How much
of the scoring could each system accept under its own review policy?
\end{enumerate}
The
first question is examined through a new comparison on matched student
records. The latter two also use the results published in the earlier
studies, with their differing confidence measures made explicit.

\section{Methodology}

\subsection{Examination, reference scores, and study labels}

The source is a written thermodynamics exam administered at ETH Zurich on
January 30, 2024. It contains four problems with 18 parts: a reactor and its
cooling system, an aircraft engine, a gas beneath a loaded piston, and a
freeze-drying process. Of 434 examinees, 252 consented to the original study,
providing 3041 pages of handwritten solutions. The original study received
ETH Zurich ethics approval 2023-N-286~\cite{Kortemeyer2024,Kortemeyer2025}.
The present analysis returns to the same collection of student work rather
than comparing different examination cohorts. In addition to considering total scores, we
also focused in on three problem parts which we selected to represent different types of open-ended problems:
\begin{itemize}
\item Analytical derivation --- deriving a symbolic expression for a physical quantity.
\item Graphical representation --- depicting a physical scenario in graphical form.
\item Numerical calculation --- calculating a numerical value for a physical quantity.
\end{itemize}
Previous studies have documented difficulties with numerical
calculations~\cite{kortemeyer2023could}, interpreting handwritten
mathematical work and recognizing equivalent
expressions~\cite{levine2026automated}, and interpreting
graphical representations~\cite{polverini2024evaluating,
polverini2024performance,kortemeyer2025multilingual,
polverini2025multimodal}. These findings motivate examining
the three types of work separately.

In the original examination, each problem was graded by two
teaching assistants, and some examinations were subsequently regraded
by a third~\cite{Kortemeyer2024}. For the present analysis, we obtained
the human reference scores from the point awards recorded in the
detailed grading sheets and summed these entries to obtain scores
for each problem part. Blank scoring cells were treated as zero,
and recorded partial credit was preserved. Where a single spreadsheet
entry covered several rubric steps, we included the recorded award
in the corresponding part total without assuming how it was divided
between those steps. For the three selected parts, the resulting
human scores agree with those in the recovered 2024 data and provide
a common reference for the comparisons across all three years.
The 2026 analysis includes 252 examinations with both TA and AI
records; confidence filtering is applied within this matched sample.
The reported statistics measure agreement with the recorded TA
scores. We did not undertake a new independent human regrading
to resolve differences between TA and AI awards.

The year labels require a chronological qualification. ``2024'' and ``2025''
refer to the publication years of the earlier studies. The first study used
tools available before the release of GPT-4o; the second used GPT-4o and
GPT-4-Turbo and was submitted in October 2024, before its publication in
April 2025~\cite{Kortemeyer2024,Kortemeyer2025}. GPT-4o itself was introduced
in May 2024~\cite{OpenAI2024GPT4o}. ``2026'' denotes the new scoring workflow.
The three labels therefore identify successive studies and their practical
systems, rather than three independently established calendar-year model
frontiers.

\subsection{Scoring Implementation}

\subsubsection{2024 Study}

The first study used a staged transcription-and-scoring
pipeline~\cite{Kortemeyer2024}. Mathpix converted the scanned handwriting
into \LaTeX. Regions that remained as images, including diagrams, were
interpreted by GPT-4V and replaced by textual descriptions. GPT-4-32k then
scored this machine-readable representation using the problem statement and
the scoring guidance. Thus a vision-capable model was part of the system,
but the final scorer worked from the resulting text. The model family and
its vision extension are described in Refs.~\cite{OpenAI2023GPT4,OpenAI2023Vision}.

The study tested four workflows with different scoring granularity and
transcription procedures. We use its workflow~2 (WF2), which covered all
252 exams and all 18 parts, for the full-exam historical comparison. It
assigned scores to problem parts using the worked solution and part maxima.
Each part score was averaged over ten successful scoring runs. Missing,
incomplete, or out-of-range outputs led to repeated requests before the
scores entered the analysis. Other workflows either covered only Problem~1
or used a smaller, 20-exam sample and are not substituted for this baseline.
For WF2, each scoring request covered one complete problem and
returned a separate score for each of its parts.

Context limits prevented the 2024 system from processing an entire
handwritten examination as a single multimodal input. In the
GPT-4V-only transcription workflow, page images therefore had to
be processed individually. Adding the problem statements could
exceed the available input capacity and could also bias the
transcription toward the solution the model expected to see.
Transcription introduced further difficulties when important
features of diagrams were lost in their textual descriptions.
A separate limitation arose during scoring: when GPT-4-32k
was asked to return 22 rubric-item scores for a complete problem,
it frequently omitted scores or returned values outside the
permitted ranges. Requesting four or five part scores instead
substantially reduced these bookkeeping
failures~\cite{Kortemeyer2024}. The practical constraints therefore
concerned both the amount of material that could be supplied and
the number of scoring decisions that could be handled reliably.
Related studies have shown that LLMs can overlook relevant
information, confuse items, or produce unsupported answers as
input contexts become longer or tasks require tracking more
information, even within the models' nominal context
windows~\cite{Liu2024LostMiddle,Hsieh2024RULER}.

\subsubsection{2025 Study}

The second study replaced the transcription stage with GPT-4o.
Individual page images were converted into \LaTeX, while diagrams
were described in German prose. The page transcriptions were combined
into a student-solution file. GPT-4-Turbo then scored one problem part
at a time. Each scoring request included the problem statement and
the transcribed student solution for all parts of the problem.
The rubric items and requested judgments were restricted to the
selected part; no separate worked sample solution was
supplied~\cite{Kortemeyer2025}.

For each rubric item, the model reported how fully the student's
work met the criterion, expressed as a percentage, and provided
a reason. We averaged these percentages over ten independent
scoring runs. Let $q_{jk}$ denote the average fulfillment for
student $j$ and criterion $k$, expressed as a fraction between
zero (not fulfilled) and one (fully fulfilled). If $w_k$ is the
maximum number of points for that criterion, the raw score is
\begin{equation}
  A_j = \sum_k w_k q_{jk}.
  \label{eq:weighted}
\end{equation}
For example, an average fulfillment of $0.6$ for a criterion
worth two points contributes $1.2$ points to the score.
We retain these fractional point awards in the raw scores.
The value $0.6$ determines how much credit is assigned; it
does not establish a $60\%$ probability that the grading
decision is correct.

The rubric was refined using repeated-run variation, item-response
analysis, and correlations between item scores. The paper reports an
initial rubric, a first refinement, and a final refinement (this accounts
for three full-inclusion points in the 2025 total-score results).
The TA scores were withheld from the rubric-refinement analyses, as
reported in the original study~\cite{Kortemeyer2025}. Although this workflow
made more extensive use of multimodal transcription, its scoring model
still evaluated a textual description of the handwritten diagrams.

\subsubsection{2026 Study}

For the 2026 analysis, we used OpenAI's GPT-6 Astra through the Azure OpenAI Responses API. The scoring program was configured with reasoning effort \texttt{high} and a maximum output budget of 60\,000 tokens per request.

\begin{figure*}
\centering
\begin{tabular}{@{}c@{}c@{}c@{}}
\small Part 1(b): Analytical derivation &
\small Part 2(a): Graphical representation &
\small Part 3(a): Numerical calculation\\[2pt]
\plotfile{0.33\textwidth}{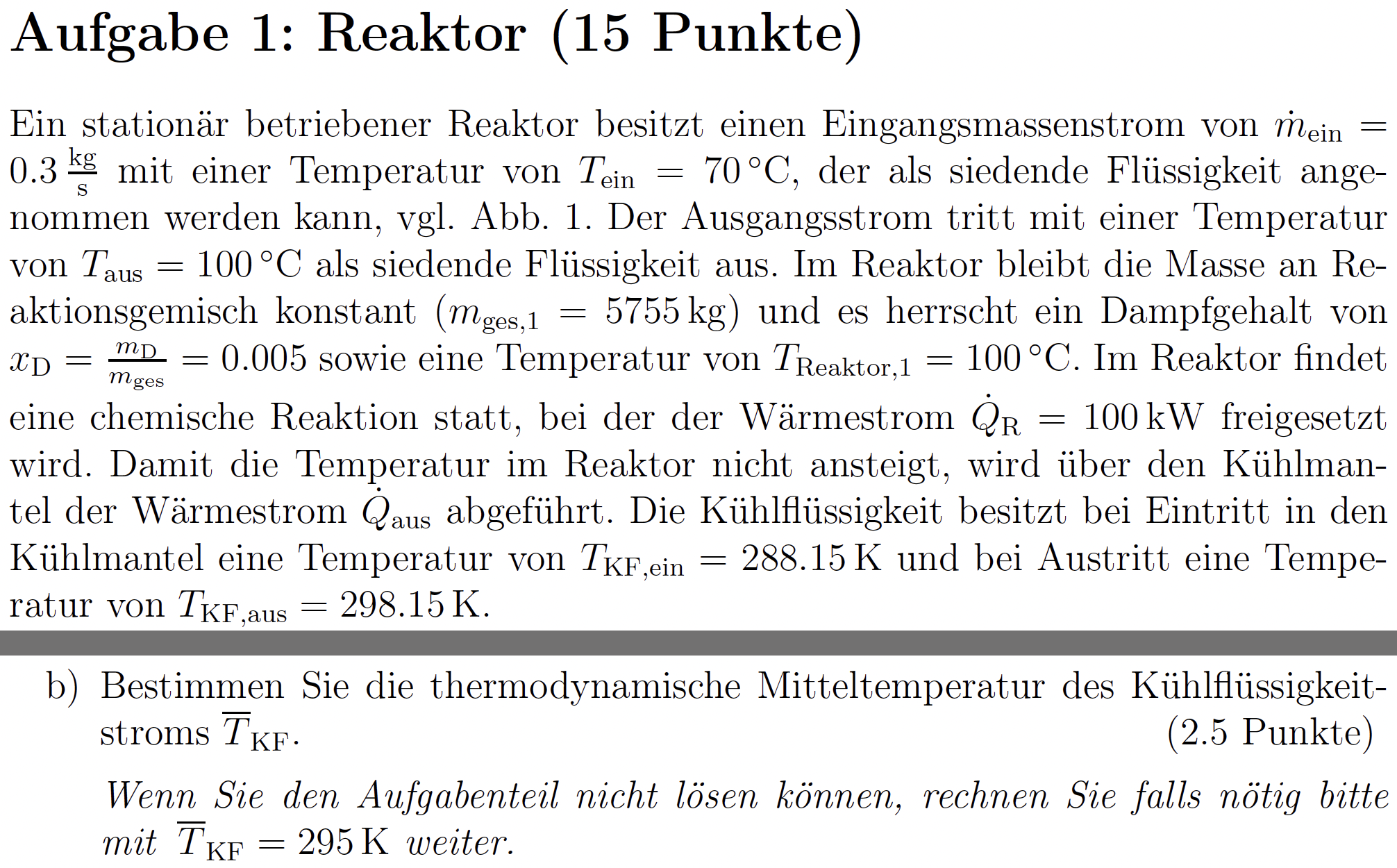} &
\plotfile{0.33\textwidth}{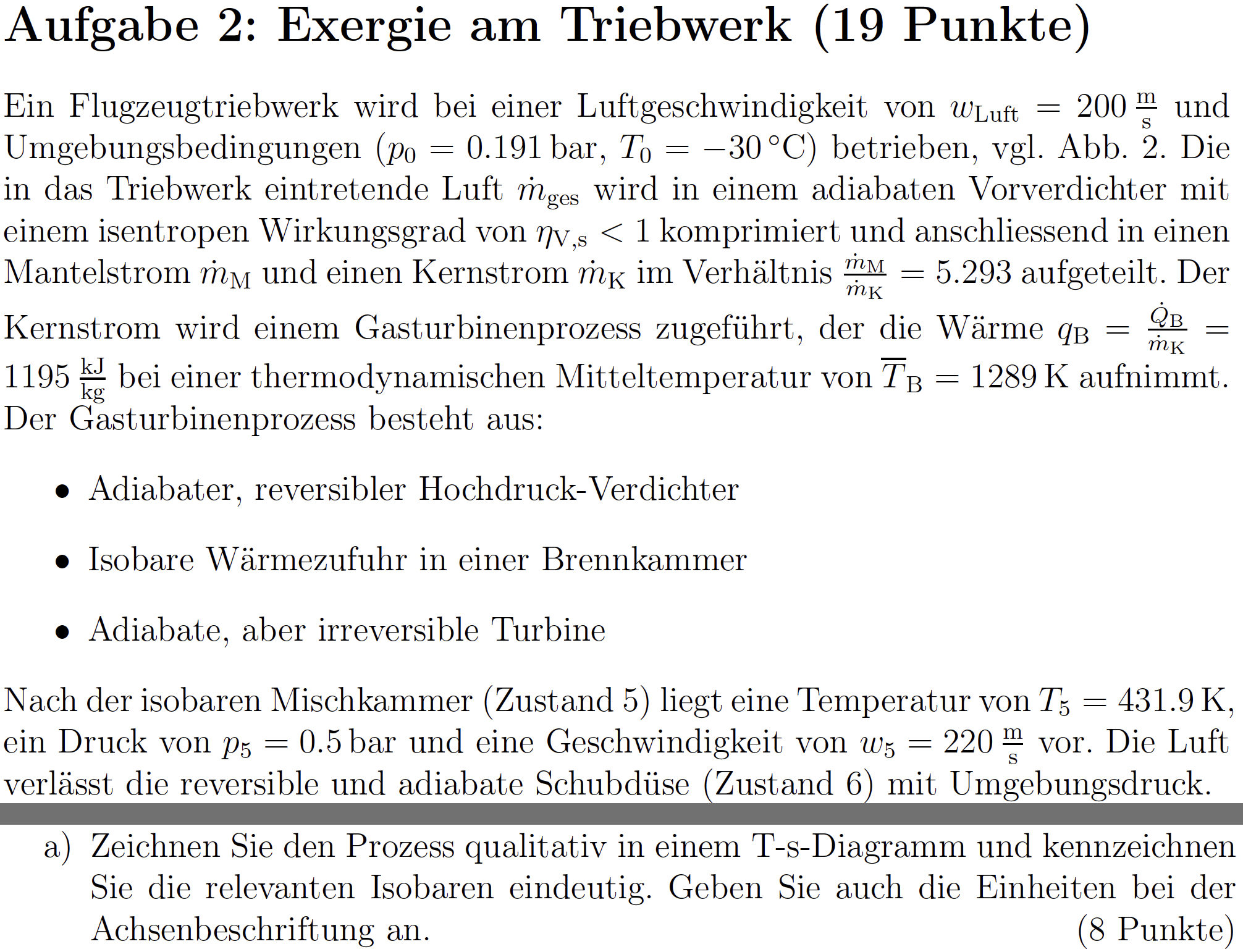} &
\plotfile{0.33\textwidth}{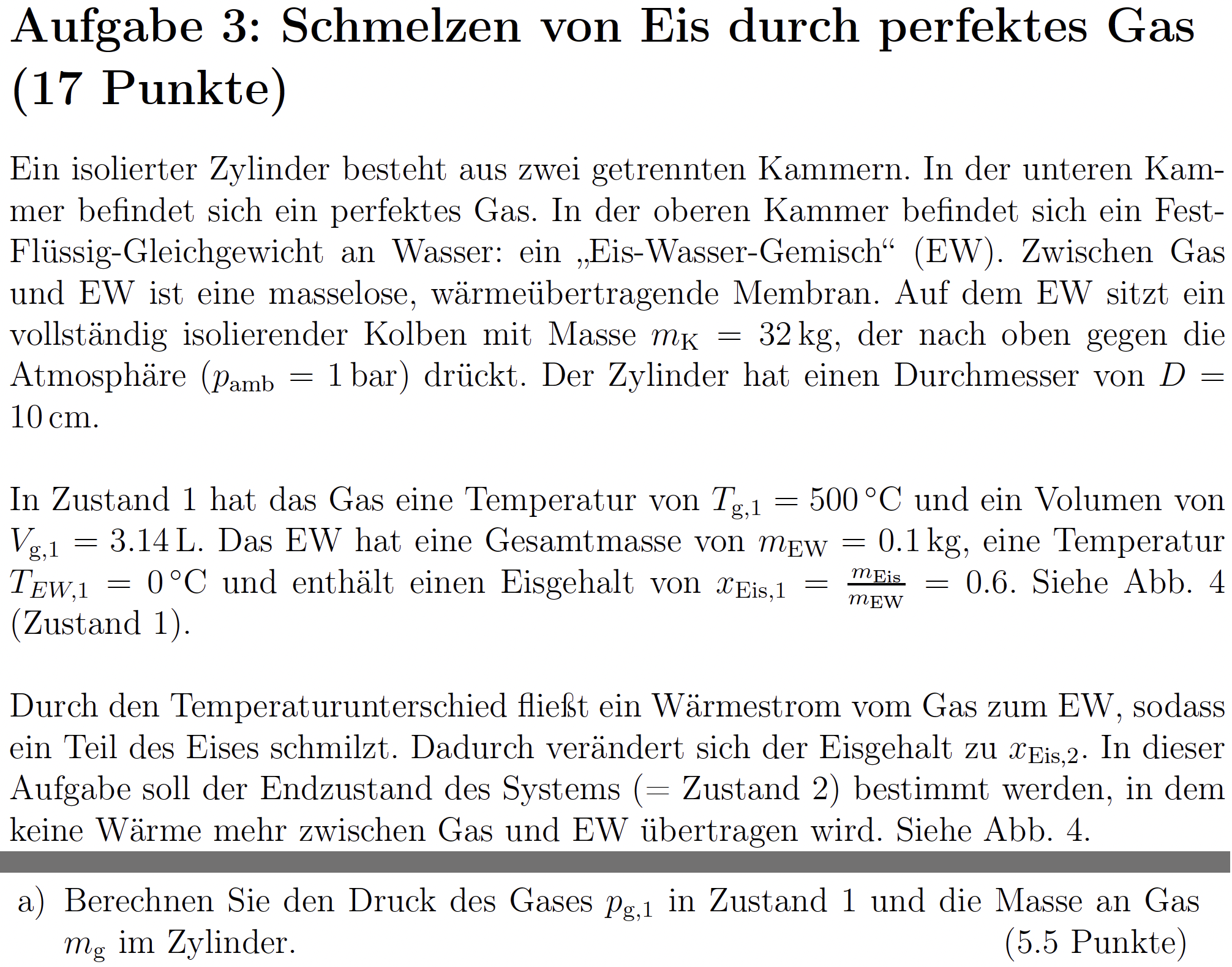}\\
\plotfile{0.33\textwidth}{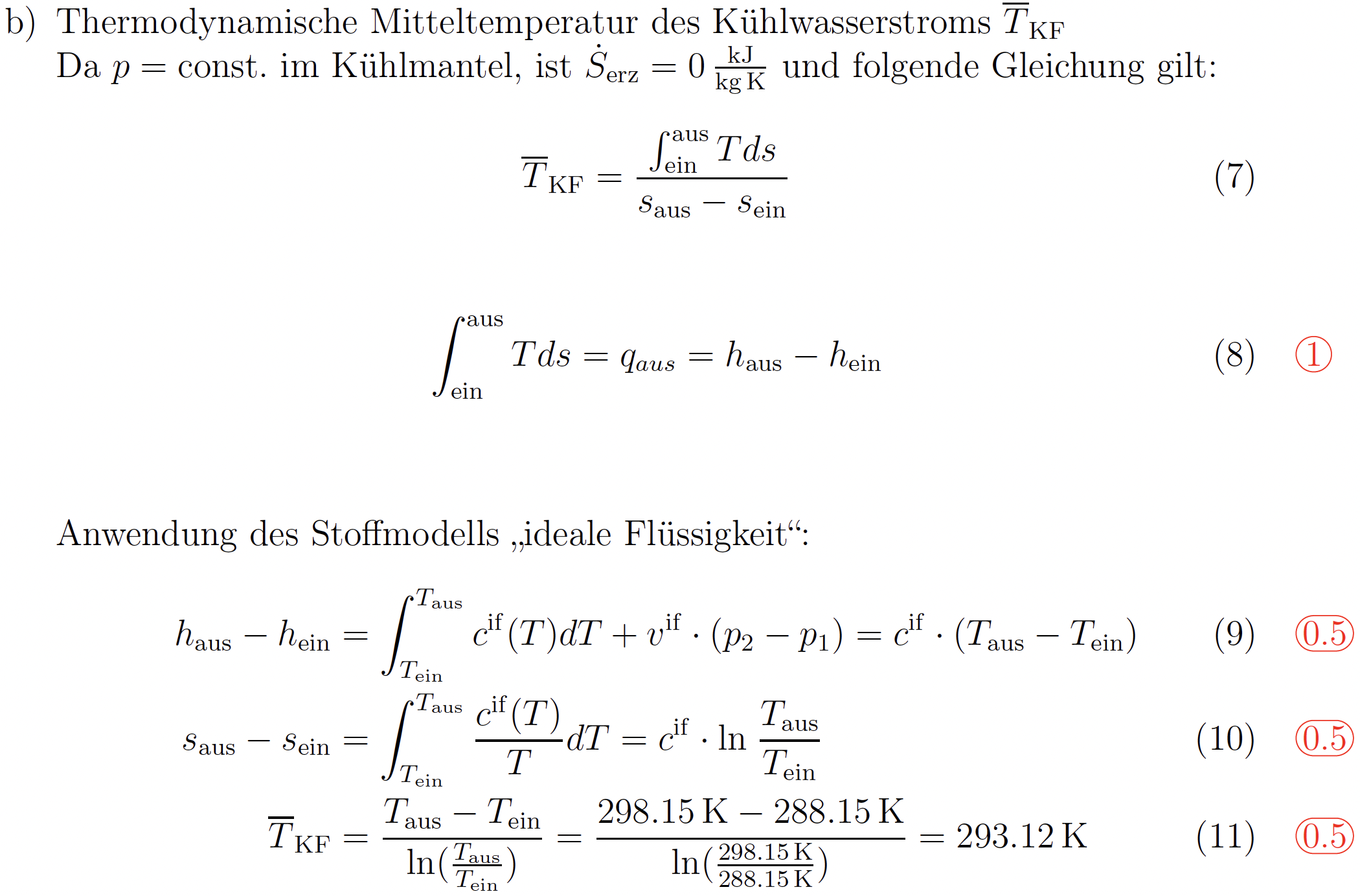} &
\plotfile{0.33\textwidth}{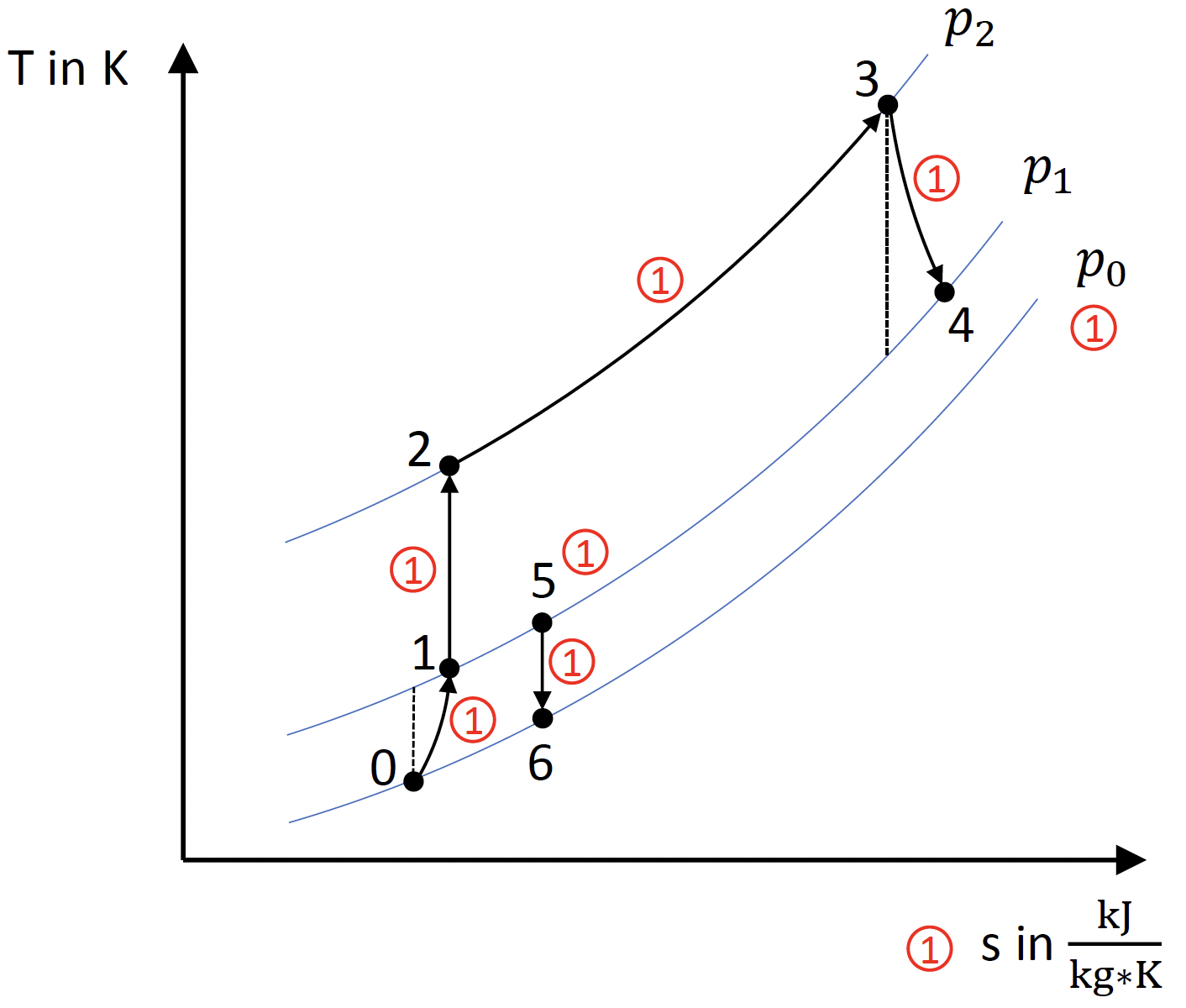} &
\plotfile{0.33\textwidth}{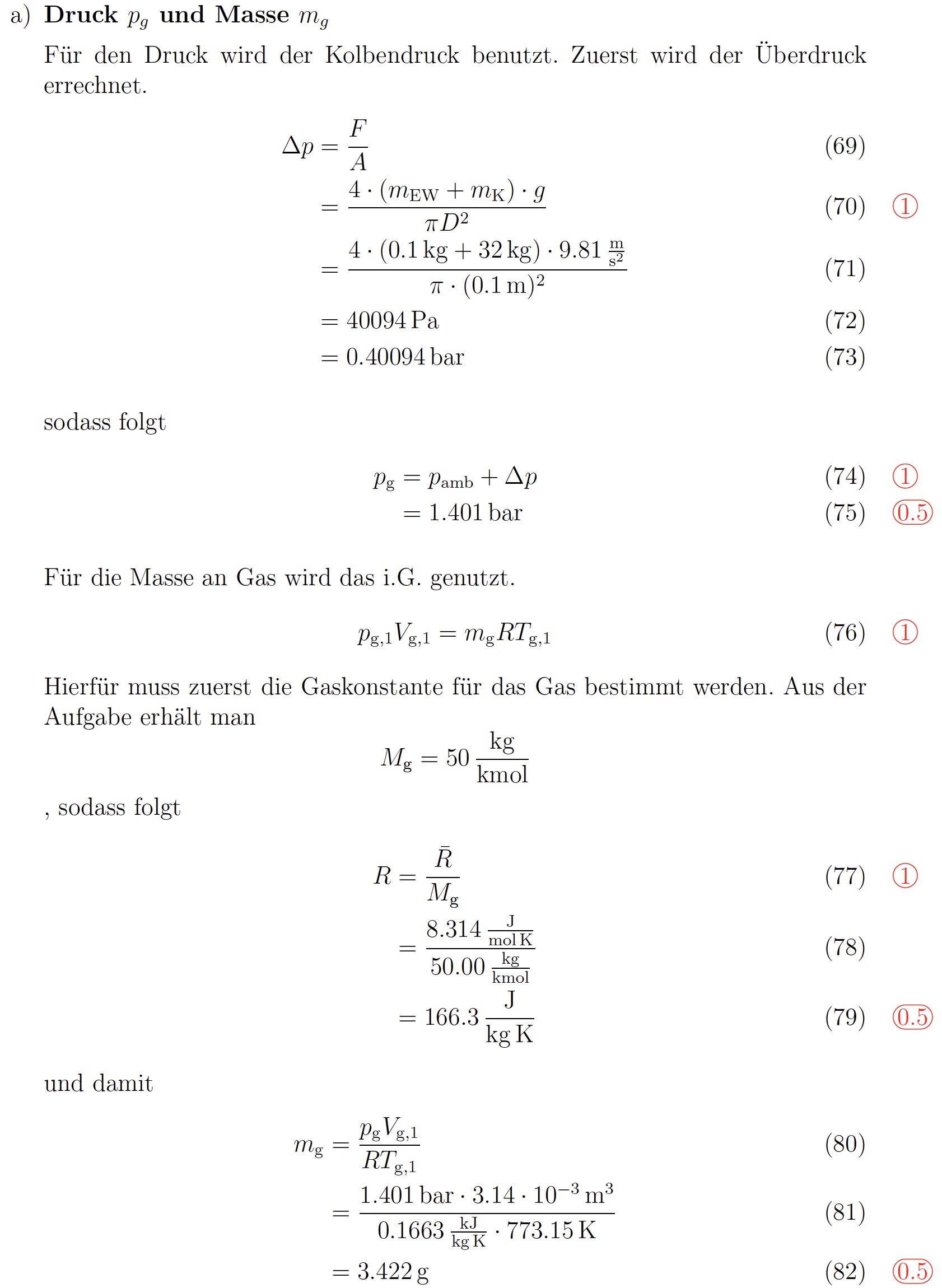}
\end{tabular}

\caption{Questions and scoring rubrics for the three selected parts.
The top row shows the original German questions; the bottom row shows
the corresponding worked solutions and scoring schemes. The numbers
in red circles specify the points awarded for individual solution
steps or features. From left to right: Part~1(b) asks students to derive
and evaluate the thermodynamic mean temperature of a coolant as it
warms while absorbing heat; this leads to a logarithmic mean of its
inlet and outlet temperatures. Part~2(a) asks for a qualitative
temperature--specific-entropy ($T$--$s$) diagram of an aircraft-engine
process, including the relevant states, isobars, and axis units.
Part~3(a) asks students to determine the initial pressure and mass of
a gas beneath a loaded piston, using mechanical equilibrium and the
ideal-gas equation. These tasks emphasize analytical derivation,
graphical representation, and numerical calculation, respectively.}
\label{fig:question_rubric_examples}
\end{figure*}

Each scoring request included the complete student examination PDF and the complete 25-page reference PDF containing the questions, figures, worked solutions, and scoring rubric. Figure~\ref{fig:question_rubric_examples} shows examples of the questions and scoring rubric; the red-circled numbers in the reference specify the points available for individual steps or features. A German scoring prompt and a machine-readable criterion catalogue, automatically generated by ChatGPT (GPT-6 Astra),  provided the expected evidence, point values, accepted alternatives, and rules for crediting consistent follow-through after an earlier error. The model assessed the original handwritten equations, calculations, and diagrams directly from the PDFs, with all examination parts available together in the same request. This allowed work continued across pages and information elsewhere in the examination to be considered when interpreting an answer. No separate transcription stage was used.

The current criterion catalogue contains 95 atomic scoring criteria in
18 parts, with a maximum of 69.5 points. Each criterion receives either
zero or its full point value. Partial credit within a part results from
summing the fulfilled criteria. The prompt accepts equivalent derivations
and consistent follow-through after an earlier error, and makes explicit
how to handle ambiguities or inconsistencies in the supplied reference.
Consequently, this is an updated operational rubric based on the same
examination, not a byte-for-byte copy of either historical prompt.

The model returns structured criterion awards, explanations, associations
between problem parts and physical PDF pages, and recommendations for human
review. The program checks that every required criterion occurs exactly
once, that awards are allowed, and that page indices and review information
are valid. It computes totals with exact rational arithmetic. These checks
prevent missing records and arithmetic inconsistencies; they do not verify
the physics of every grading decision.

One completed scoring response is retained per exam. The implementation
does not form an average over independent grading repetitions. Retries
after technical failures are distinct from the ten-run averages used in
the earlier studies. Full-document input and a single structured scoring
response thus represent a different use of the available multimodal and
context capabilities. The comparison measures the complete resulting
system, including these changes in input and aggregation.

\subsection{Confidence Measures}

All three workflows can leave work for a human grader, but they use different
evidence to decide when to do so. In machine learning, such an option is
often called selective prediction: a system accepts some decisions and
abstains on others~\cite{Geifman2017}. Here, ``accepted'' means retained
under the specified analysis rule. It does not establish that an individual
award is correct, and it is unrelated to a student later accepting or contesting
a final examination score.

\subsubsection{2024 Study}

The confidence measure was the dispersion of the repeated scores. A small
standard deviation indicated that the scoring procedure reproduced its
own award. The published full-exam selection required a summed standard
deviation below 10 points and an AI total above one point, the latter
excluding very low totals associated with transcription failure.
It retained 212 of 252 exams, or 84.1\%~\cite{Kortemeyer2024}.
This was a whole-exam selection: retained exams contributed their complete
scores. As it turned out, a repeated error in the transcription could nevertheless yield a
small score dispersion, so repeatability was not sufficient evidence of
correctness.

\subsubsection{2025 Study}

The second study compared each mean AI judgment with the pattern expected
from a two-parameter logistic item-response model. In that model,
\begin{equation}
  p_{jk}=\frac{1}{1+\exp[-\alpha_k(\theta_j-\beta_k)]},
  \label{eq:irt}
\end{equation}
where $\theta_j$ is a student parameter, $\beta_k$ is item difficulty, and
$\alpha_k$ is item discrimination. The parameters were estimated from the
AI-scored response patterns. The uncertainty measure was
\begin{equation}
  u_{jk}=|p_{jk}-q_{jk}|.
\end{equation}
It described disagreement with the psychometric pattern, rather than a
model's explicit estimate of its own confidence~\cite{Kortemeyer2025}.

Three controls determined acceptance, where ``unexpected'' AI-scoring results could be set aside for human review. A correctness threshold $C$ converted
mean fulfillment to full or zero credit, an uncertainty threshold $U$
limited the discrepancy above, and an optional positive-only policy $P$
retained only judgments classified as correct. The same accepted
student--criterion pairs were then summed in the AI and TA scores.
Unlike the raw scores in Eq.~\eqref{eq:weighted}, these filtered AI scores
used thresholded, binary awards. Selection therefore changed both the
included work and the credit rule.

The historical confidence analysis also excluded parts for which the AI
awarded zero on every item, treating them as presumed skipped work.
This convention matters: the original paper reports that 43\% of those
parts nevertheless received some TA credit~\cite{Kortemeyer2025}.
An all-zero AI score cannot therefore be used as an independent annotation
that the student did no work.

\subsubsection{2026 Study}

The current system produces qualitative review recommendations scoped to
problem parts. Reasons include material illegibility, conflicting final
answers, missing or uncertain page evidence, and consequential ambiguities
in applying the rubric. Flagging such items was left to the reasoning-capable LLM; the exact criteria applied by the model to identify such items were thus opaque. Structural checks in the scoring software could also
add a referral. A confidently identified incorrect or blank answer does
not require referral solely because it receives zero credit.

If any concern applies to a criterion, the entire associated part is
excluded from the filtered analysis. Global or unassignable concerns
exclude the complete exam. This conservative rule avoids retaining some
criteria from a part whose interpretation requires human attention.
Each excluded part is removed from both the AI and TA scores. The retained
scores are summed without rescaling them to the original examination
maximum. An exam contributes to a filtered regression if at least one
part remains within that regression's scope. Due to the opaque nature of the model decisions, the flags provide one
operating point, not a calibrated numerical confidence scale that can be
swept through many thresholds. 

\subsection{Coverage as a count of scoring judgments}

Our operational interest is how much scoring work can be accepted from
the AI. We count rubric judgments rather than awarded or available points.
This gives equal weight to checking a half-point item and checking a
one-point item. It remains a measure of scoring decisions, not a direct
measurement of minutes of human effort, e.g., it is not directly translatable into time-savings.

We measure coverage as the proportion of rubric-item judgments
accepted without a recommendation for human review. Applying one
rubric item to one student's work counts as one judgment,
regardless of the number of points awarded. To express coverage
in a common unit across the studies, we use the final historical
rubric as the counting reference. It contains 85 items per
examination: 23, 20, 19, and 23 for Problems~1--4, respectively.
In the 2026 analysis, acceptance or referral for human review
applies to entire problem parts. Let $h_p$ be the number of
historical rubric items in part $p$, and let $m_{jp}$ equal one
if that part is accepted for student $j$, and zero if it is
referred for review. For $N$ matched examinations before
filtering, coverage is
\begin{equation}
  \mathcal{C}=100\%
  \frac{\sum_{j=1}^{N}\sum_p h_p m_{jp}}
       {N\sum_p h_p}.
  \label{eq:coverage}
\end{equation}
The numerator counts accepted judgments, while the denominator
counts all judgments in the matched sample. For example, a part
containing four historical rubric items contributes four
accepted judgments if retained and none if referred for review;
in either case, all four remain in the denominator. A score of
zero does not itself trigger exclusion. The historical item
counts are used only to calculate coverage. The 2026 AI scores
continue to use the current 95-criterion rubric, and the point
totals used in the regressions are unchanged.

All parts of the 252 valid matched exams remain in the denominator,
including zero-score parts and parts removed by review flags. The overall
denominator is therefore $252\times85=21420$ judgments. ``100\% coverage''
denotes the no-abstention benchmark: all raw scores are included, including
scores the system recommended reviewing. It does not mean that the review
policy approved every score for use without a human.

The 2024 whole-exam selection translates exactly to this counting
convention: each retained exam contributes the same 85 reference judgments,
so its coverage remains $212/252$. The 2025 percentages also count
student--rubric-item judgments, but its presumed-skipped-part exclusions
and lack of published integer acceptance counts prevent an exact
reconstruction of the eligible denominator. We therefore retain its
published percentages and identify them as historical judgment coverage.
The plots share a meaningful counting unit, but identical historical
eligibility is not established. They should not be used to infer an exact
cross-year difference in human time saved or an equal-coverage accuracy
comparison.

\subsection{Matched comparisons}

\begin{figure*}
\centering
\begin{tabular}{@{}c@{}c@{}c@{}}
\small Part 1(b): Analytical derivation &
\small Part 2(a): Graphical representation &
\small Part 3(a): Numerical calculation\\[2pt]
\plotfile{0.33\textwidth}{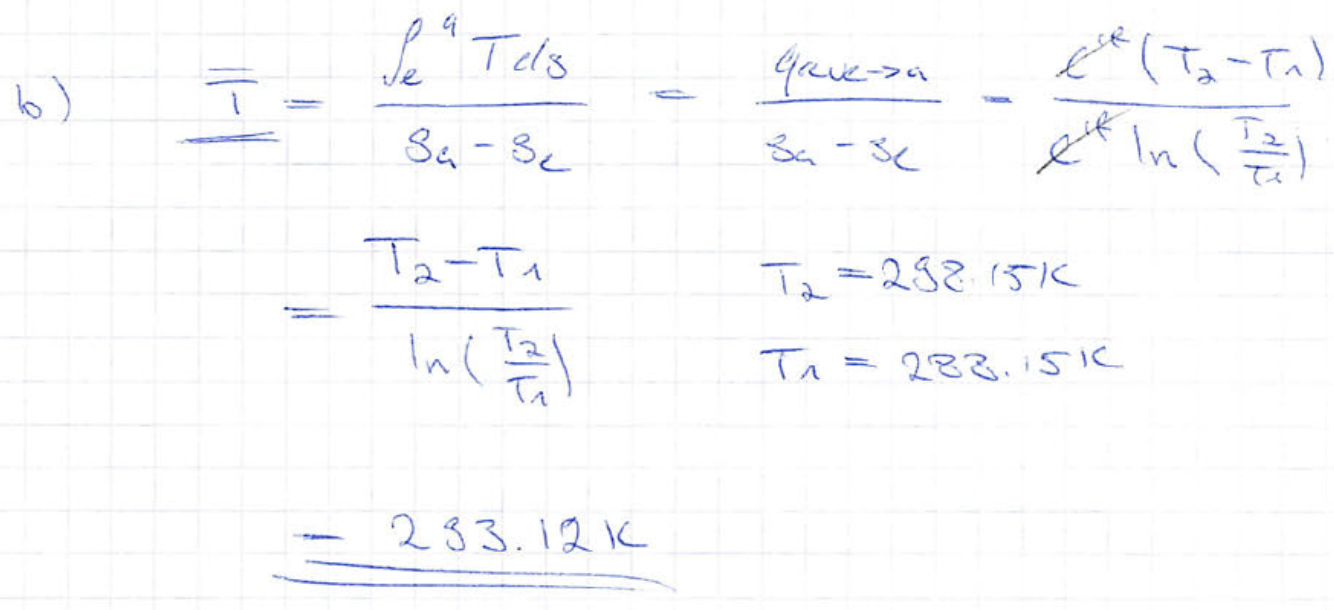} &
\plotfile{0.33\textwidth}{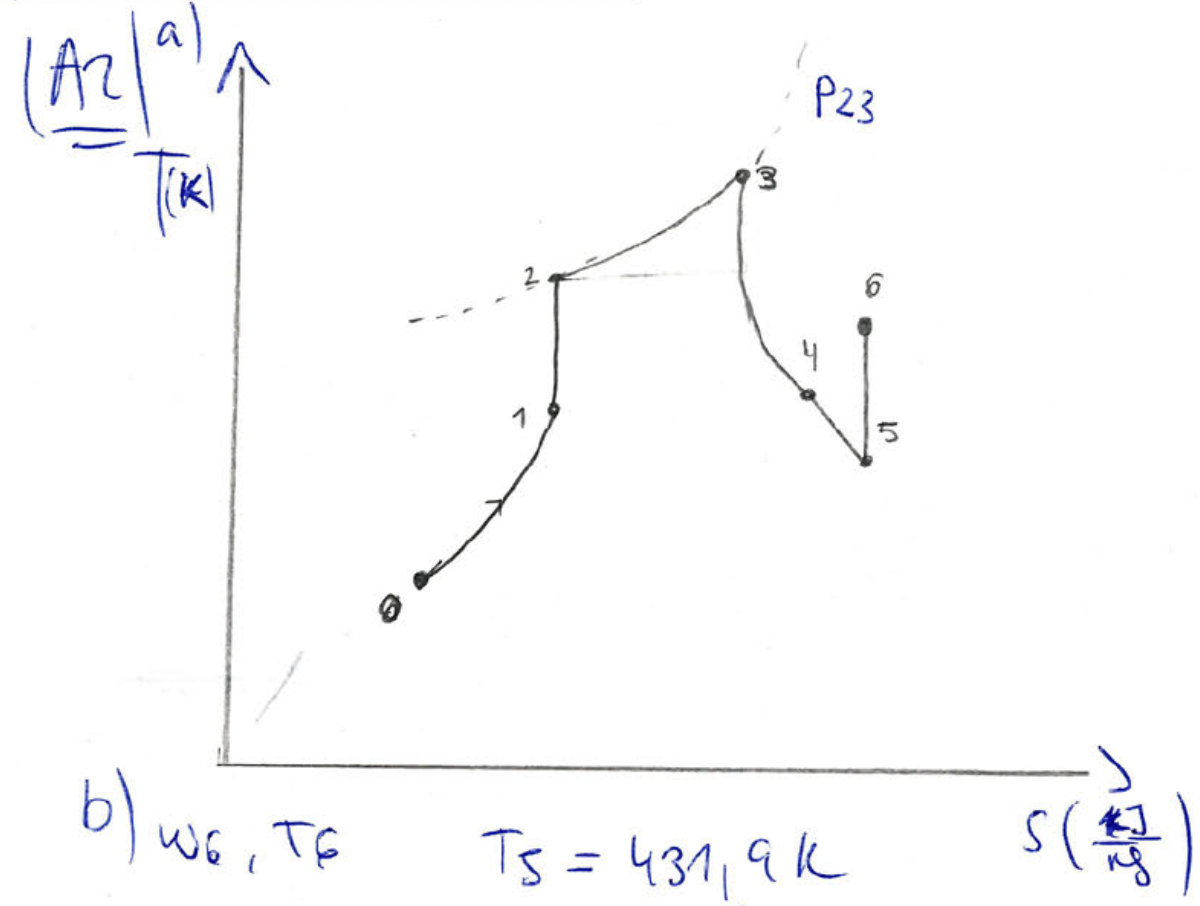} &
\plotfile{0.33\textwidth}{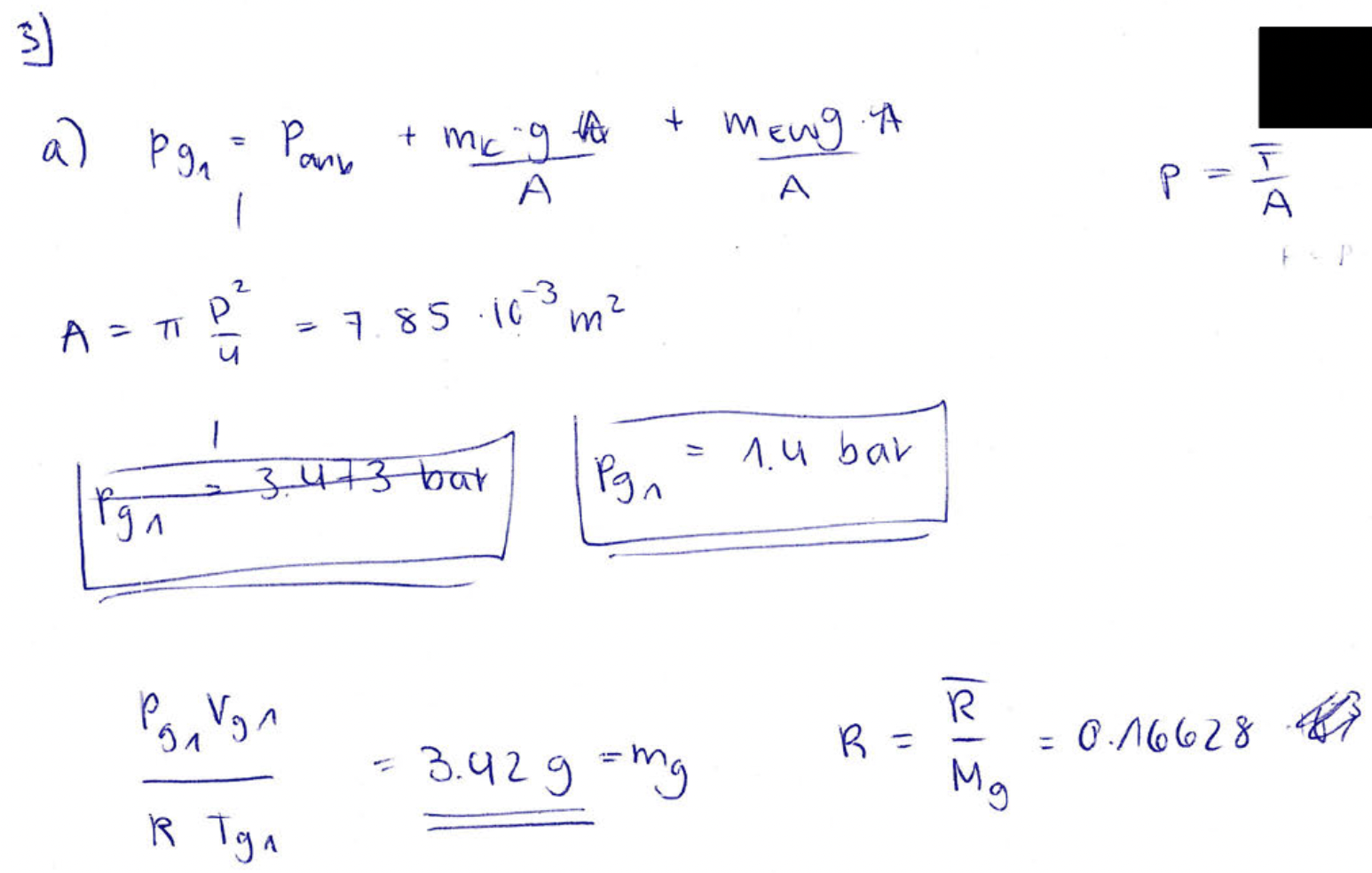}\\
\plotfile{0.33\textwidth}{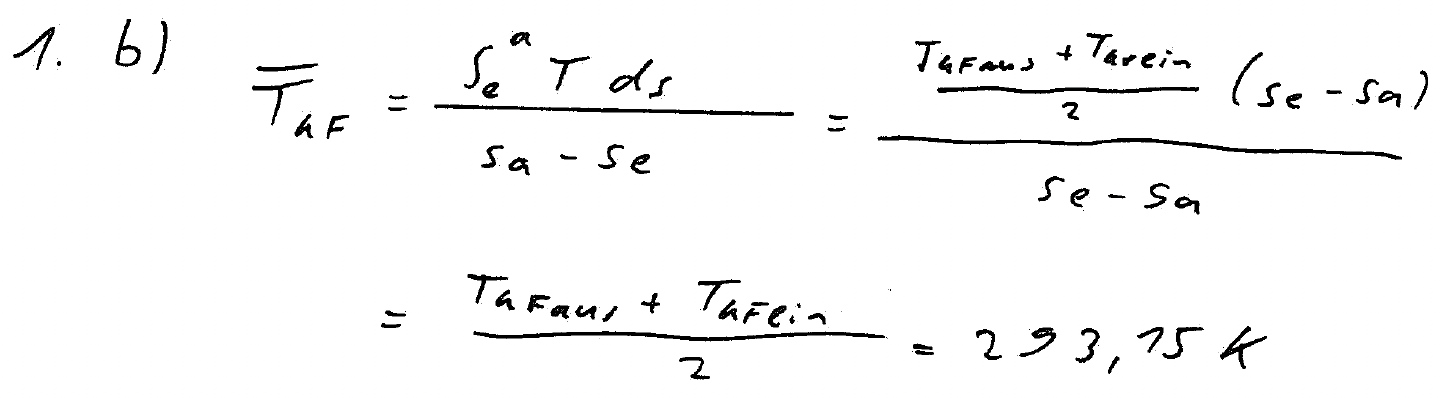} &
\plotfile{0.33\textwidth}{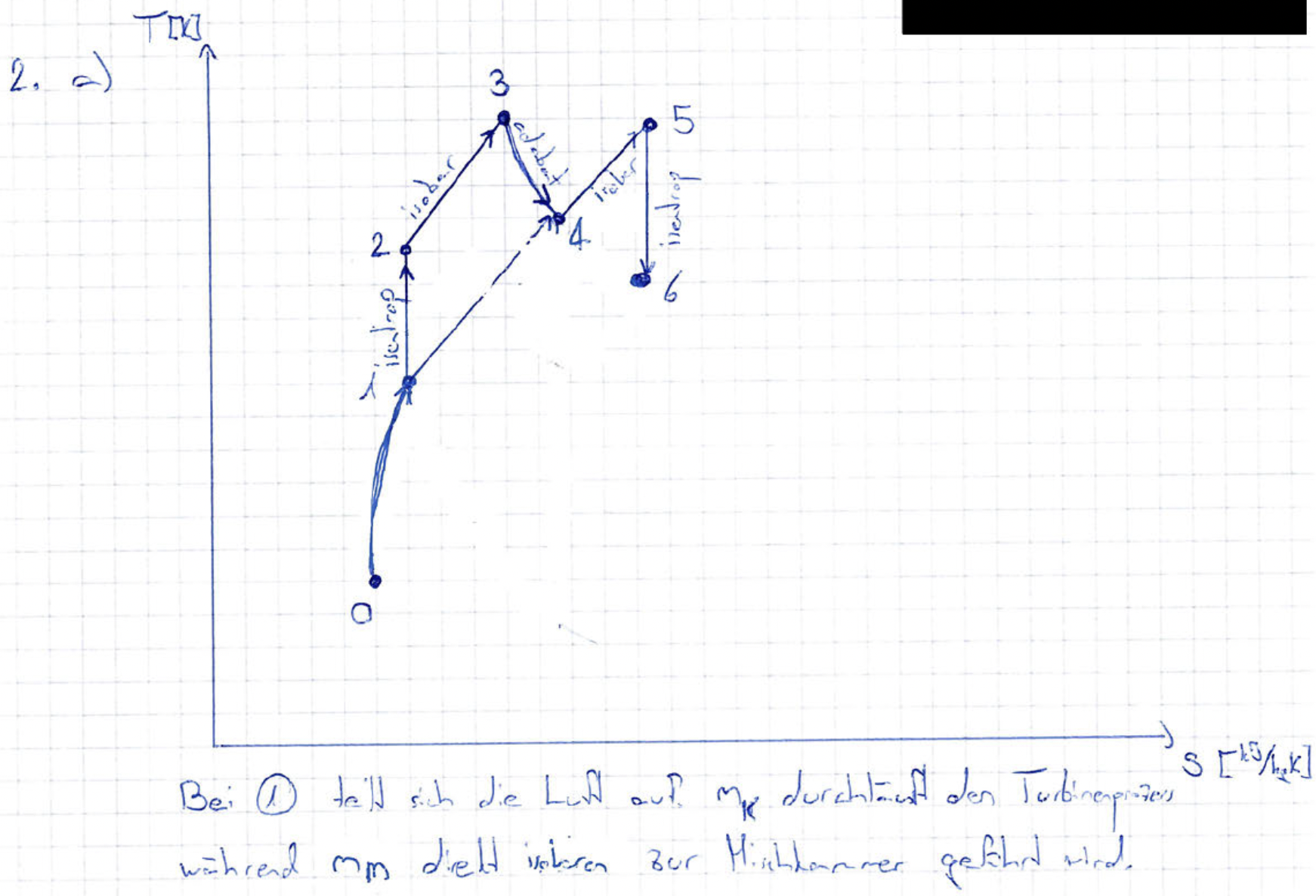} &
\plotfile{0.33\textwidth}{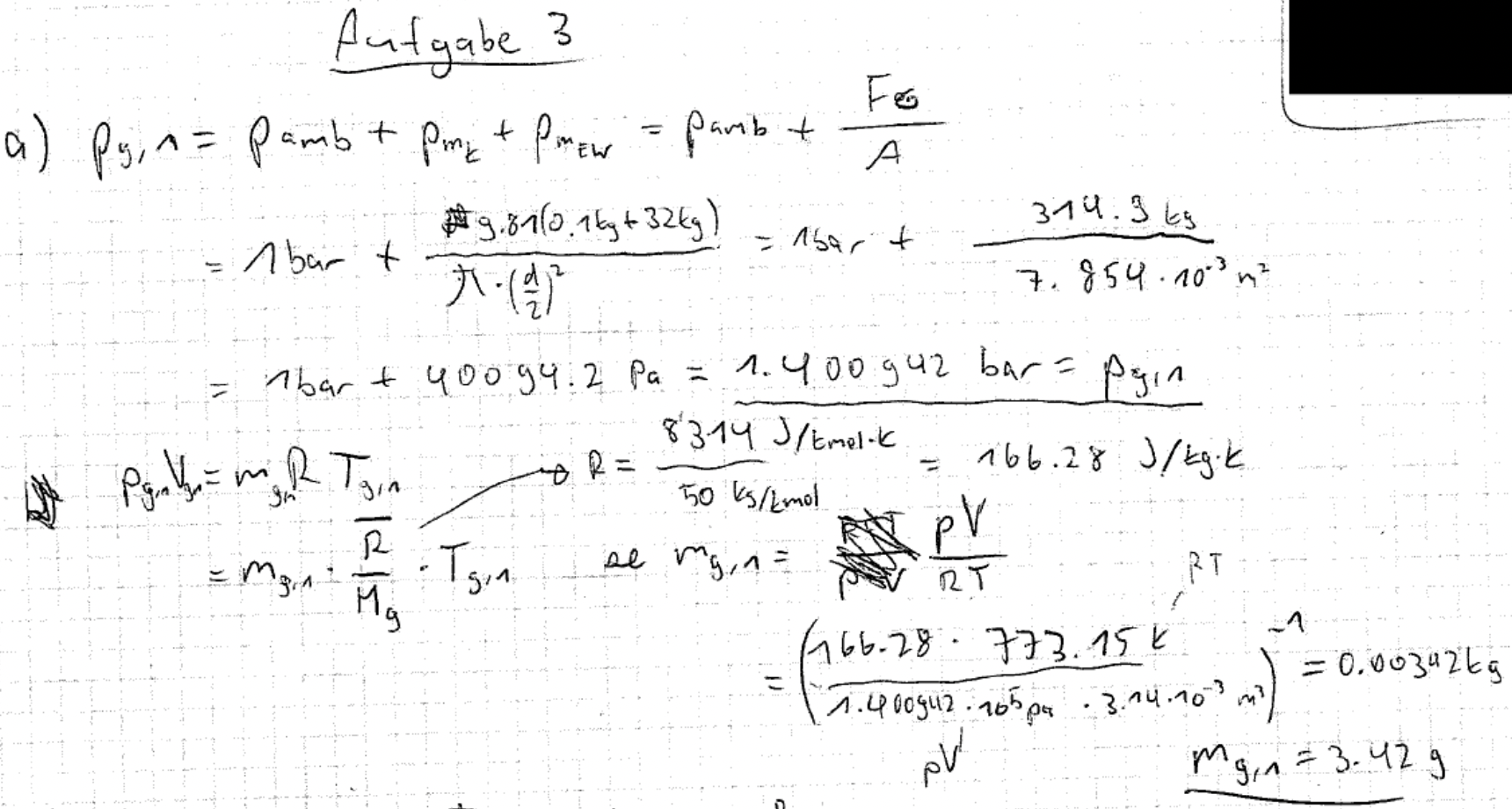}\\
\plotfile{0.33\textwidth}{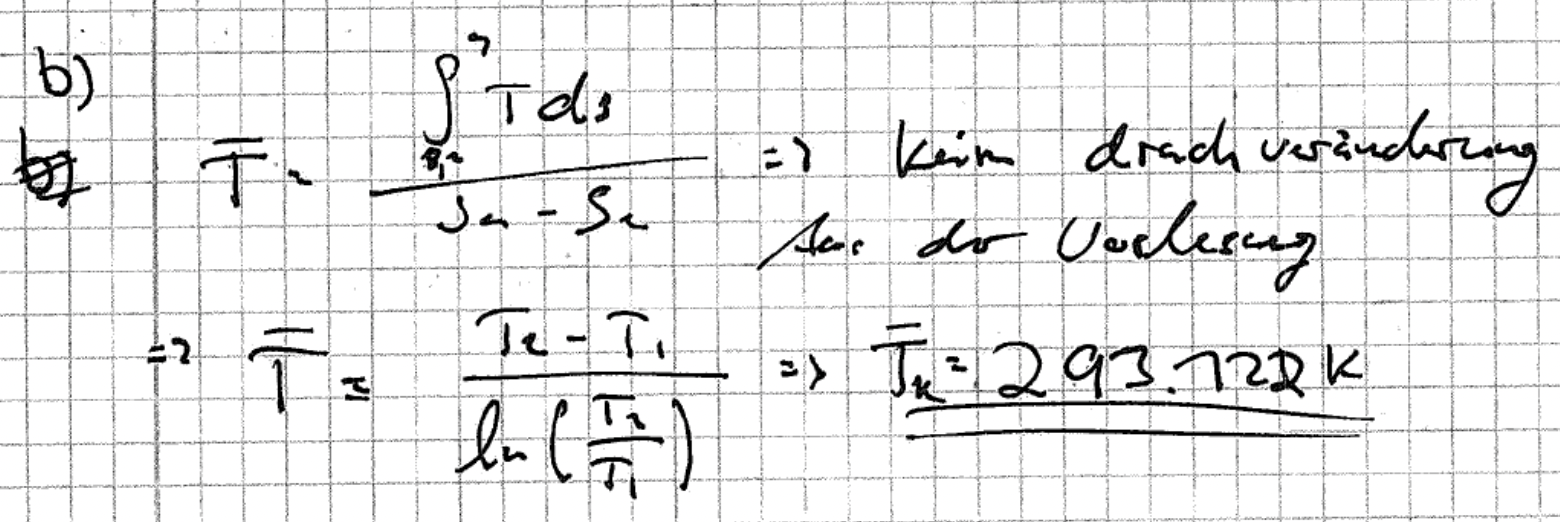} &
\plotfile{0.33\textwidth}{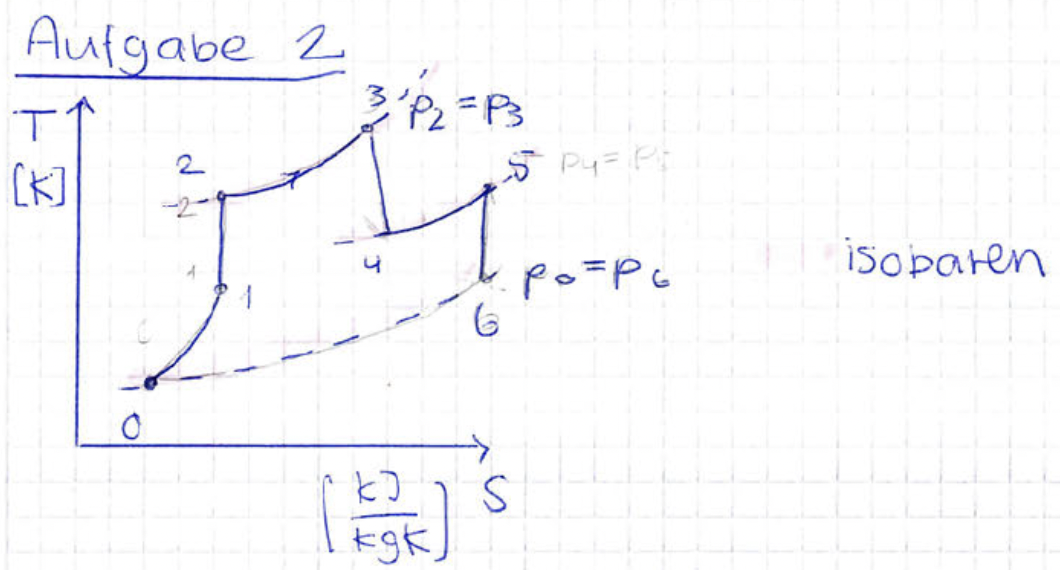} &
\plotfile{0.33\textwidth}{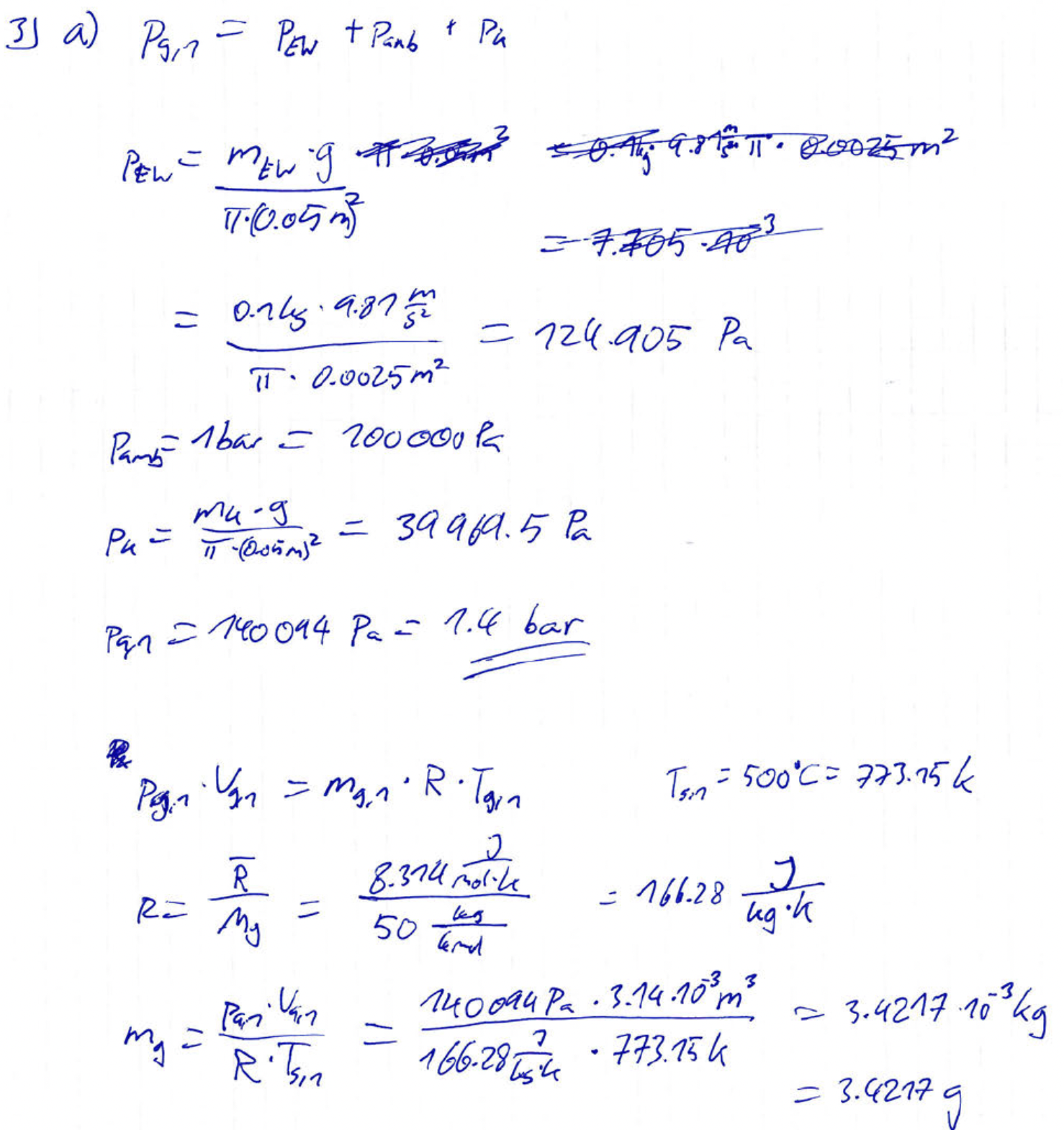}
\end{tabular}

\caption{Illustrative excerpts of anonymized handwritten student
solutions to the three selected parts: the thermodynamic mean-temperature
derivation in Part~1(b), the aircraft-engine $T$--$s$ diagram in
Part~2(a), and the calculation of gas pressure and mass in Part~3(a).
The excerpts show the types of written evidence the scoring systems
must interpret, including equations, intermediate steps, diagrams,
and numerical results. They serve as illustrations; all valid matched
student records enter the quantitative comparisons.}
\label{fig:student_examples}
\end{figure*}

We selected three contrasting parts to represent different problem types. Part~1(b), worth 2.5 points, asks for
the thermodynamic mean temperature of the coolant and requires the
logarithmic-mean relation. It represents analytical manipulation, although
it also has a numerical result. Part~2(a), worth 8 points, asks for the
qualitative aircraft-engine process in a $T$--$s$ diagram, including the
relevant isobars and axis units. Part~3(a), worth 5.5 points, asks for the
initial gas pressure and mass using mechanical equilibrium and the
ideal-gas equation. These are illustrative examples of different
representations, not three experimentally isolated categories of reasoning.
Figure~\ref{fig:question_rubric_examples} shows the exam problem parts and associated scoring rubrics under consideration; Figure~\ref{fig:student_examples} shows examples of student work on these problem parts.

The selected-part comparison uses unfiltered AI scores from all
three years. No confidence filters or human-review exclusions
are applied. The archived 2024 averages and the 2026 scores
are already expressed in points. The archived 2025 averaged
fulfillment percentages are converted to points using
Eq.~\eqref{eq:weighted}, retaining fractional
awards.\footnote{For Part~3(a), we assign the combined
2025 gas-constant criterion a weight of 1.5 points:
one point for the expression and half a point for its
numerical evaluation in the original examination rubric.}

For each selected part, we compare the same examinations across
all three years against the same TA scores. We include only
examinations with available scores in every year. This gives
252 examinations for Parts~1(b) and~2(a), and 251 for
Part~3(a), for which one 2024 record is missing. All scores
in these matched samples, including zeros, are retained,
giving 100\% coverage in each year.

\subsection{Linear regression and interpretation}

We use linear regression to describe the relationship between
AI and TA scores for selected problem parts, complete problems,
and grand totals. The same regression measures are used for
unfiltered scores and for scores retained after filtering.
The regressions are descriptive: we report fitted coefficients
and $R^2$, without hypothesis tests or confidence intervals.

The regression model is
\begin{equation}
  A_j=a+bT_j+\epsilon_j,
  \label{eq:regression}
\end{equation}
where $T_j$ and $A_j$ are the TA and AI scores for examination
$j$, and $\epsilon_j$ is the deviation from the fitted line.
Depending on the comparison, these scores refer to a problem
part, a complete problem, the whole examination, or the subtotal
for accepted work. Each newly calculated fit uses ordinary
least squares with an intercept and gives equal weight to
each examination. The offset $a$ is measured in points, and
the slope $b$ is dimensionless. We report
\begin{equation}
  R^2=1-\frac{\sum_j(A_j-\widehat{A}_j)^2}
                  {\sum_j(A_j-\overline{A})^2},
  \label{eq:r2}
\end{equation}
where $\widehat{A}_j=a+bT_j$ is the fitted AI score and
$\overline{A}$ is the mean AI score in the comparison.
Perfect agreement with the TA scores would give $a=0$,
$b=1$, and $R^2=1$.

The offset $a$ is the fitted AI score when the TA score is
zero. When the slope is one, this offset represents a constant
difference between the fitted AI scores and the TA scores.
The slope $b$ describes the change in the fitted AI score
for a one-point increase in the TA score. A slope of one
preserves the point scale; a positive slope below one
compresses the fitted score range, while a slope above one
expands it. The coefficient of determination $R^2$ gives
the fraction of variation in AI scores accounted for by
the linear fit. A value of one means that all observations
lie on the fitted line.

A large $R^2$ alone therefore does not establish agreement:
a system that always awards half the TA score has $R^2=1$
but $b=0.5$. Agreement requires both a fitted line close to
the identity line and little scatter around it. Likewise,
the offset alone is not the mean scoring bias; the fitted
AI--TA difference at TA score $T$ is $a+(b-1)T$.

The selected-part regressions are newly calculated for all
three years using the common examination samples described
above. The grand-total overview combines the published
2024 and 2025 coefficients with newly calculated 2026
results, and includes both unfiltered and filtered scenarios.
For 2024, filtering selected whole examinations and retained
their complete scores. For 2025 and 2026, the filtered
comparisons use TA and AI subtotals for the same accepted
work. The historical grand-total results therefore use
the same regression measures but do not constitute the
same strictly matched comparison as the selected-part
analysis. In particular, the historical grand-total fits
have not all been recalculated against the current TA
reference.\footnote{The 2026 TA totals are calculated from
the detailed grading sheets. Seven detailed-sheet totals in the full source workbook
differ from its summary totals.
The historical total-score reference vectors have not
all been reconstructed to check these discrepancies.}

\section{Results}

\subsection{Raw scores for three problem parts}

\begin{figure*}
\centering
\begin{tabular}{@{}c@{}c@{}c@{}}
\small Part 1(b): Analytical derivation &
\small Part 2(a): Graphical representation &
\small Part 3(a): Numerical calculation\\[2pt]
\plotfile{0.33\textwidth}{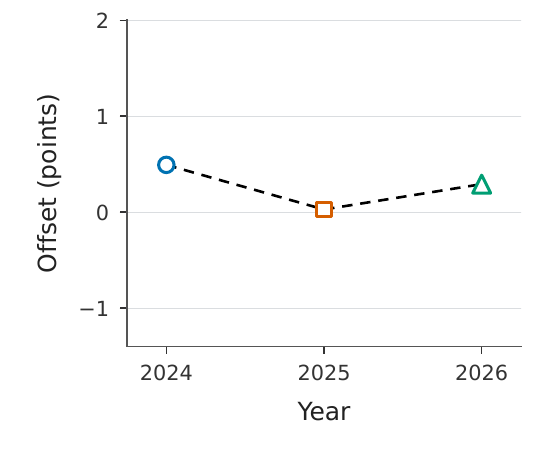} &
\plotfile{0.33\textwidth}{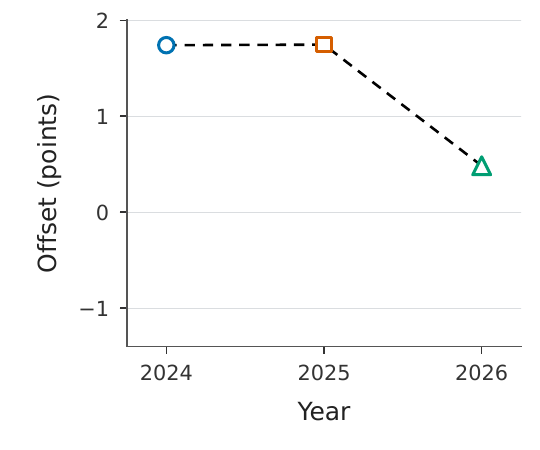} &
\plotfile{0.33\textwidth}{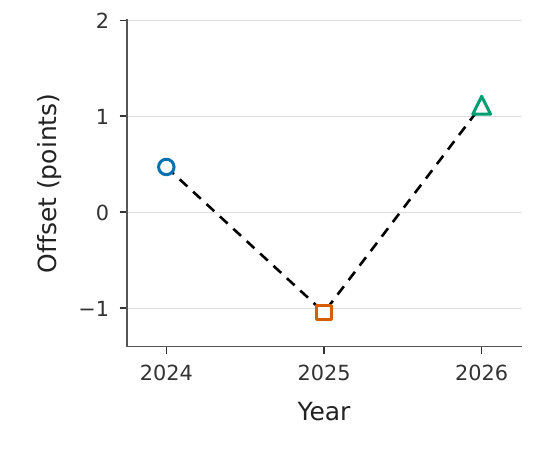}\\
\plotfile{0.33\textwidth}{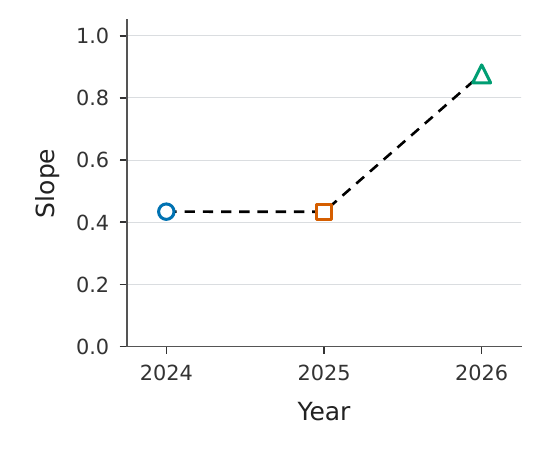} &
\plotfile{0.33\textwidth}{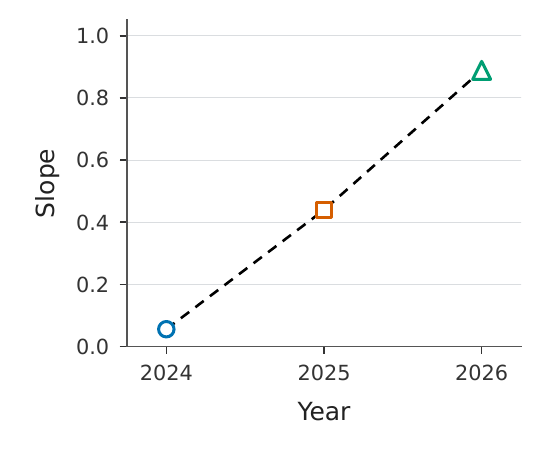} &
\plotfile{0.33\textwidth}{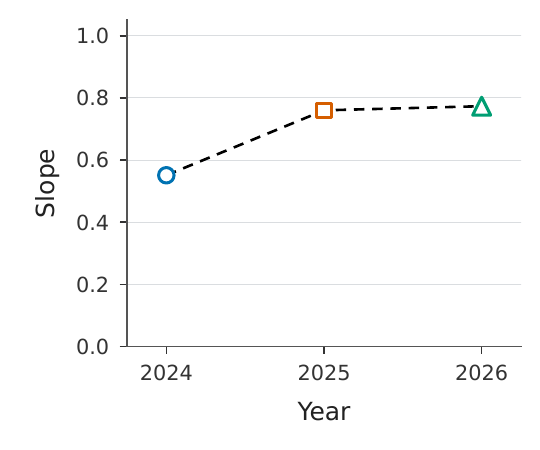}\\
\plotfile{0.33\textwidth}{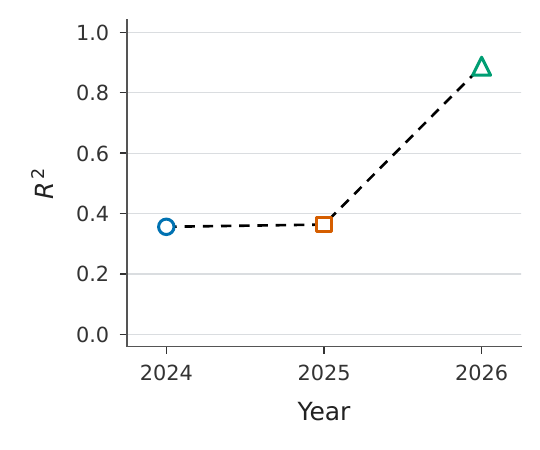} &
\plotfile{0.33\textwidth}{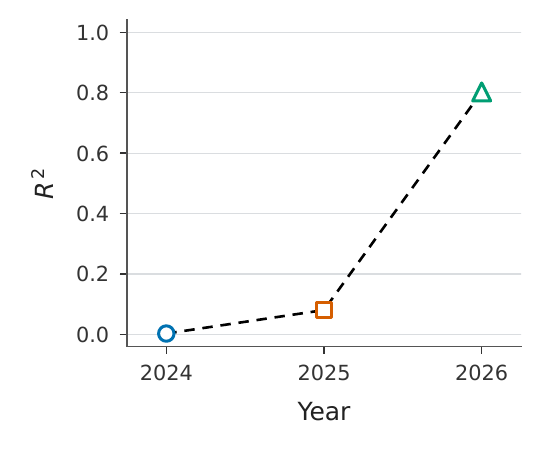} &
\plotfile{0.33\textwidth}{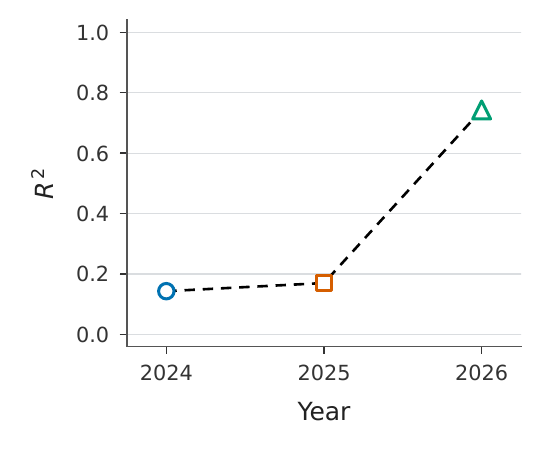}
\end{tabular}
\caption{Raw-score regression coefficients for the three selected problem
parts at 100\% coverage. Columns show Parts~1(b), 2(a), and 3(a); rows
show offset, slope, and $R^2$. Blue circles, orange squares, and green
triangles denote the 2024, 2025, and 2026 studies. Black dashed lines connect
the successive studies as guides to the eye. The cohort is held fixed
across years within each part: $N=252$, 252, and 251, respectively.
The natural agreement values are zero offset and unit slope; $R^2$
describes scatter about the fitted line.}
\label{fig:problemtypes}
\end{figure*}

Table~\ref{tab:parts} and Fig.~\ref{fig:problemtypes} compare
the three years using the same student responses for each
problem part. All scores for these responses are included,
without confidence filtering or exclusions for recommended
human review (100\% coverage).

\begin{table*}
\caption{Raw-score comparison on the same exams within each part, with
100\% coverage in every row. For Part~3(a), one historical scoring record was unavailable in the
2024 data; the comparison therefore uses the same 251 exams in all
three years. The regression is AI score on TA score, with an intercept. Offsets are in examination points. Coefficients are rounded to
four decimal places.}
\label{tab:parts}
\begin{ruledtabular}
\begin{tabular}{llrrrrr}
Part & Main representation & Year & $N$ & Offset & Slope & $R^2$\\
\hline
1(b) & Analytical derivation & 2024 & 252 & 0.4930 & 0.4340 & 0.3562\\
     &                       & 2025 & 252 & 0.0278 & 0.4335 & 0.3637\\
     &                       & 2026 & 252 & 0.2920 & 0.8774 & 0.8879\\
\addlinespace
2(a) & Graphical representation & 2024 & 252 & 1.7402 & 0.0556 & 0.0029\\
     &                          & 2025 & 252 & 1.7457 & 0.4395 & 0.0806\\
     &                          & 2026 & 252 & 0.4835 & 0.8888 & 0.8027\\
\addlinespace
3(a) & Numerical calculation & 2024 & 251 & 0.4708 & 0.5511 & 0.1431\\
     &                       & 2025 & 251 & -1.0452 & 0.7601 & 0.1701\\
     &                       & 2026 & 251 & 1.1144 & 0.7734 & 0.7431\\
\end{tabular}
\end{ruledtabular}
\end{table*}

For the analytical derivation in Part~1(b), the earlier fits are similar:
the slopes are approximately 0.43 and $R^2$ is approximately 0.36 in
both years. In 2026 the slope increases to 0.877 and $R^2$ to 0.888.
The offset remains positive, at 0.292 points on a 2.5-point part.
Thus the improvement concerns both the scaling of the scores and their
association with the TA awards, without complete agreement with the
identity line.

The largest change in $R^2$ occurs for the graphical task in Part~2(a).
In 2024 the slope is 0.056 and $R^2=0.003$, indicating almost no linear
association between the AI and TA scores. The 2025 workflow improves the
slope to 0.440, but $R^2$ remains only 0.081. In 2026 the slope is 0.889
and $R^2=0.803$, while the offset decreases from about 1.74 to 0.48
points. The improvement is therefore not merely a shift in the mean
level of the award.

For the numerical calculation in Part~3(a), $R^2$ increases from 0.143
and 0.170 to 0.743. The slope, however, changes little between 2025 and
2026, from 0.760 to 0.773, while the offset changes from $-1.045$ to
$1.114$ points. This result is a useful reminder that the three metrics
describe different features. The newer scores follow the variation in
the TA scores more closely, but their fitted scale and offset still
differ from perfect agreement. Progress is not monotonic in every
coefficient.

\subsection{Grand totals and accepted scoring coverage}
\begin{figure*}
\centering
\plotfile{0.49\textwidth}{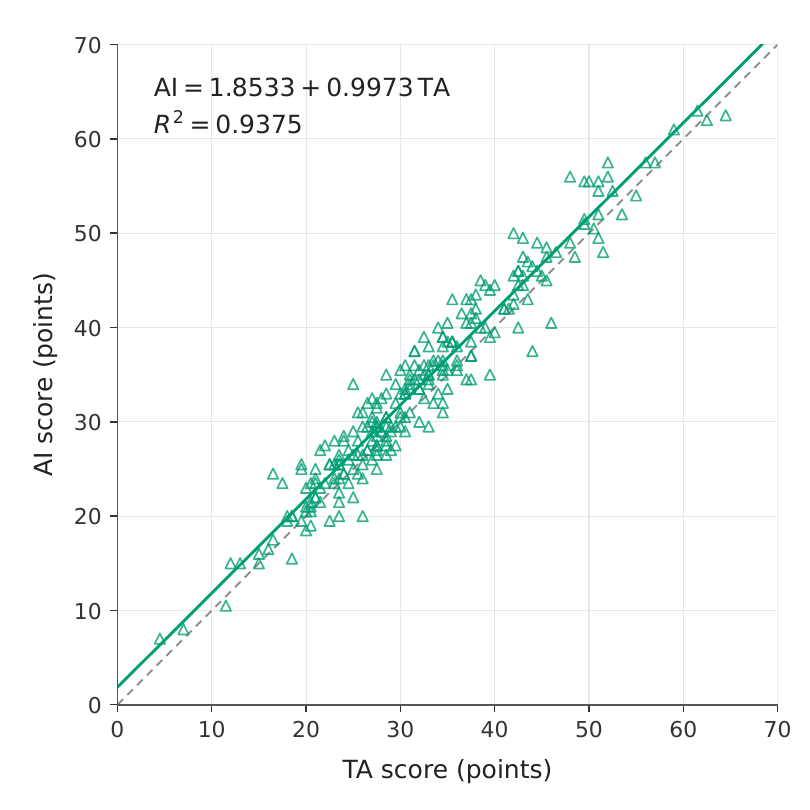} 
\plotfile{0.49\textwidth}{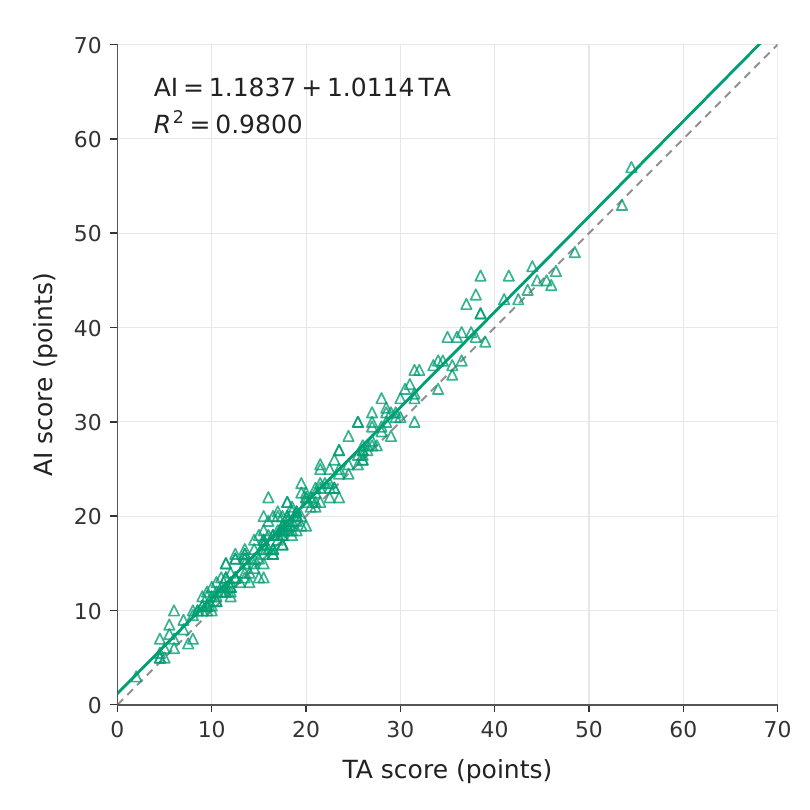} 
\caption{TA and AI scores for the 2026 analysis. Left: complete-exam
totals at 100\% coverage ($N=252$). Right: subtotals after excluding
review-flagged problem parts from both the TA and AI scores
(62.8\% coverage, $N=250$). Solid green lines show the least-squares
fits; dashed gray lines indicate equal TA and AI scores.
While the scatter plot makes it appear like scores were lowered, this is simply due to fewer total points being available after removing the flagged problem parts from consideration --- the filtered scores are subtotals for retained work and
are not rescaled to the original examination maximum.}
\label{fig:scatter}
\end{figure*}

Figure~\ref{fig:scatter} compares complete-exam totals with retained subtotals after excluding review-flagged problem parts from both the TA and AI scores in the 2026 analysis.

\begin{figure}
\centering
\plotfile{0.92\columnwidth}{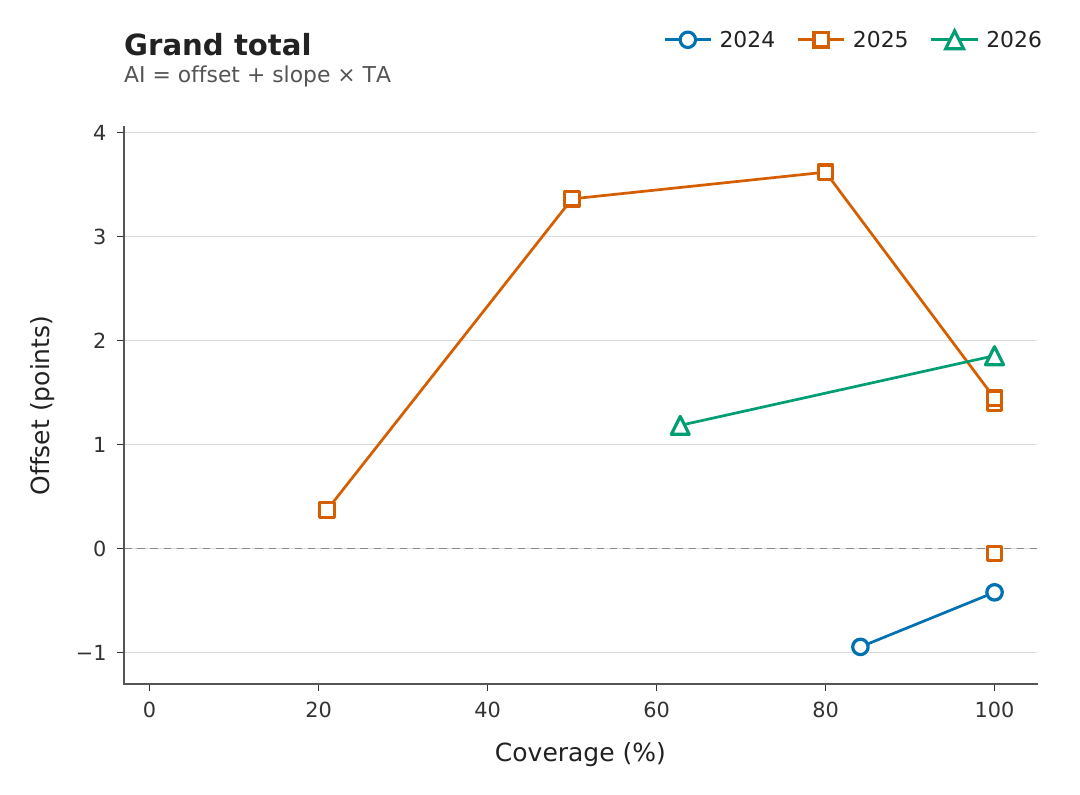}\par
\plotfile{0.92\columnwidth}{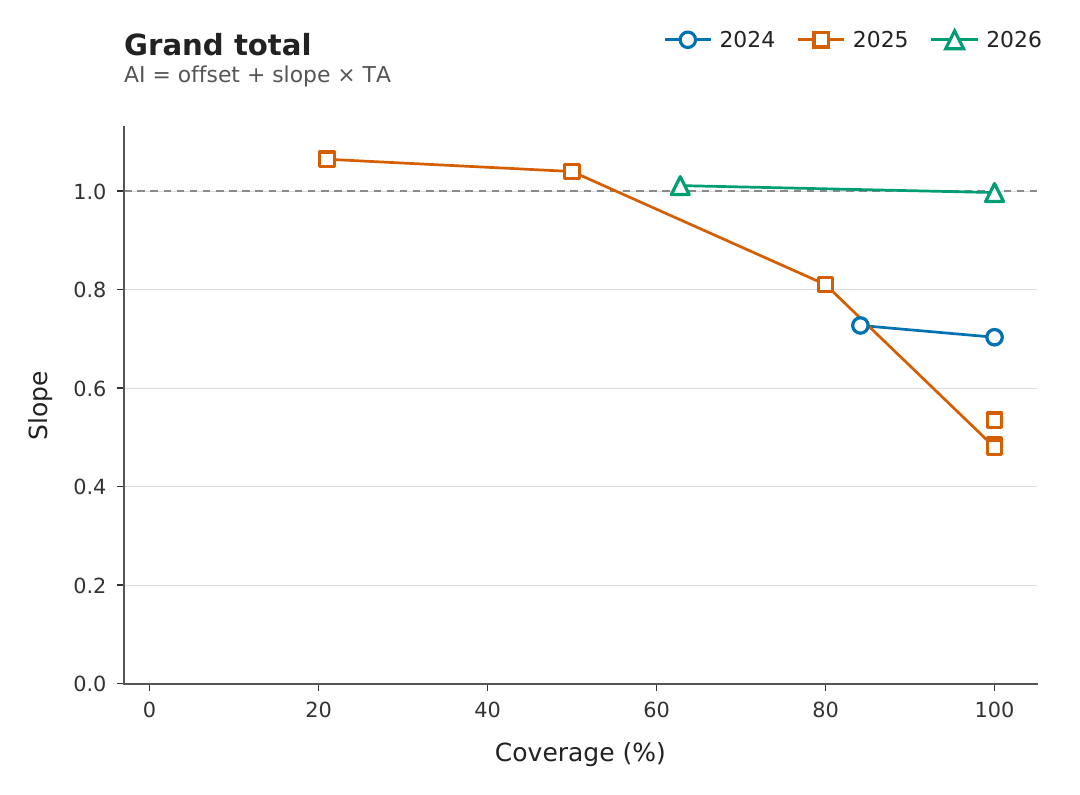}\par
\plotfile{0.92\columnwidth}{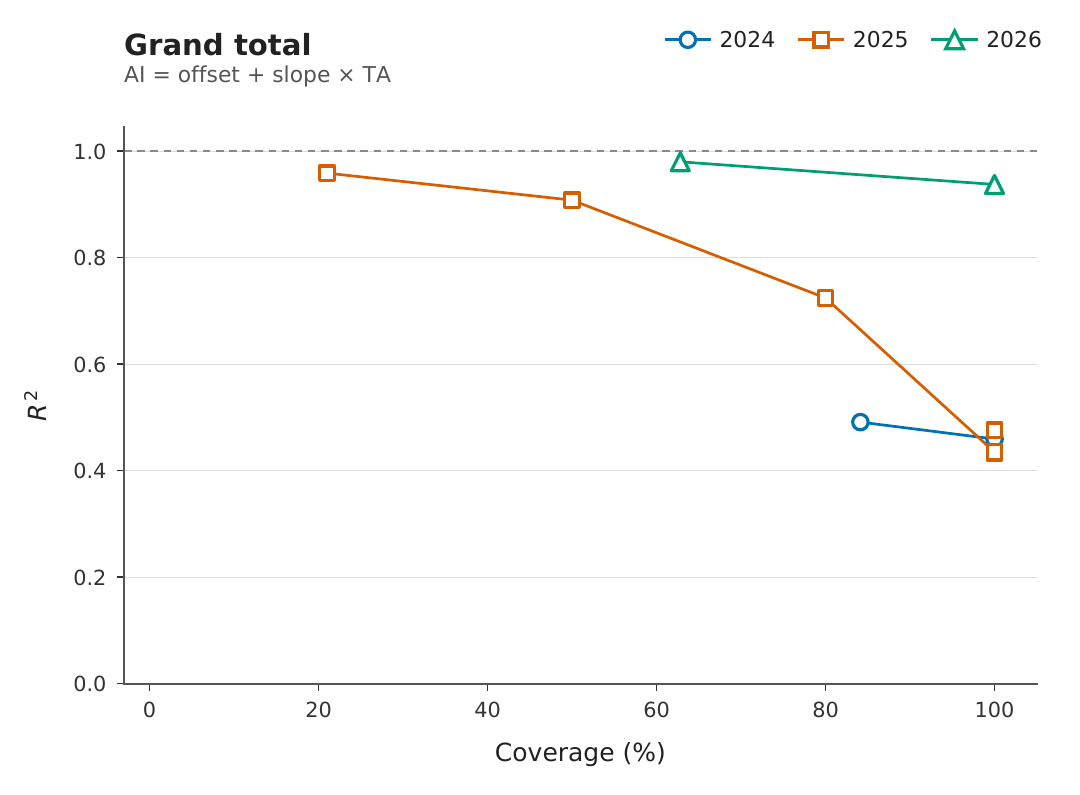}
\caption{Grand-total offset, slope, and $R^2$ versus accepted judgment
coverage. Blue circles denote 2024, orange squares 2025, and green
triangles 2026. The three 2025 markers at 100\% represent the initial,
first-refinement, and final-refinement rubrics; the line uses the final
endpoint. Lines are guides to the eye. Coverage counts rubric judgments,
with the published eligibility convention retained for 2025. Filtered
fits concern different selected work and, in 2025, binary rather than
continuous AI awards; see Methods.}
\label{fig:totals}
\end{figure}

Table~\ref{tab:totals} and Fig.~\ref{fig:totals} place the complete-exam
results alongside the historical operating points. The unfiltered 2024
WF2 result has a slope of 0.704 and $R^2=0.459$. The three unfiltered
2025 rubric stages have slopes from 0.480 to 0.535 and $R^2$ from 0.434
to 0.476. The 2026 raw-total fit has a slope of 0.997 and $R^2=0.937$,
with an offset of 1.853 points. Its nearly unit slope therefore coexists
with a positive score shift.

\begin{table*}
\caption{Grand-total regression results. Historical coefficients and
2025 coverage percentages are reproduced from
Refs.~\cite{Kortemeyer2024,Kortemeyer2025}; 2026 values come from the
current analysis. ``All'' denotes no abstention. The filtered 2024 fit
uses complete scores for selected exams; the filtered 2025 and 2026
fits use accepted subtotals. The selected 2025 coverage percentages
follow the original study's rules for excluding parts judged unanswered,
as described in Methods. The transcription and scoring columns identify
the tools and models used at each stage. Scoring context denotes the student work available in a single scoring
request. The 2024 and 2025 scorers received transcribed work for all
parts of the problem; the 2026 scorer received the complete student
examination PDF together with the complete reference PDF. In 2025,
the rubric and requested judgments were restricted to the selected
part despite the broader student-work context. Confidence granularity denotes the unit accepted or rejected by
confidence filtering. A dash indicates that no confidence filtering
was applied. The merged 2026 model entry
indicates direct PDF scoring by GPT-6 Astra without a separate
transcription stage.}
\label{tab:totals}
\begingroup
\footnotesize
\setlength{\tabcolsep}{2pt}
\begin{ruledtabular}
\begin{tabular}{llccccrrrr}
Study & Configuration &
\shortstack{Transcription\\model/tool} &
\shortstack{Scoring\\model} &
\shortstack{Scoring\\context} &
\shortstack{Confidence\\granularity} &
Coverage (\%) & Offset & Slope & $R^2$\\
\hline
2024 & WF2, all &
\shortstack{Mathpix\\+ GPT-4V} & GPT-4-32k &
\shortstack{Problem\\(text)} & -- &
100.0 & -0.4218 & 0.7035 & 0.4594\\

2024 & WF2, selected whole exams &
\shortstack{Mathpix\\+ GPT-4V} & GPT-4-32k &
\shortstack{Problem\\(text)} & Whole exam &
84.1 & -0.9462 & 0.7274 & 0.4912\\

2025 & Initial rubric, all &
GPT-4o & GPT-4-Turbo &
\shortstack{Problem\\(text)} & -- &
100.0 & -0.0471 & 0.5350 & 0.4759\\

2025 & First refinement, all &
GPT-4o & GPT-4-Turbo &
\shortstack{Problem\\(text)} & -- &
100.0 & 1.3992 & 0.4847 & 0.4337\\

2025 & Final refinement, all &
GPT-4o & GPT-4-Turbo &
\shortstack{Problem\\(text)} & -- &
100.0 & 1.4443 & 0.4800 & 0.4349\\

2025 & $C=0.1$, $U=0.5$, $P=\mathrm{false}$ &
GPT-4o & GPT-4-Turbo &
\shortstack{Problem\\(text)} & Rubric item &
80 & 3.6177 & 0.8104 & 0.7243\\

2025 & $C=0.1$, $U=0.5$, $P=\mathrm{true}$ &
GPT-4o & GPT-4-Turbo &
\shortstack{Problem\\(text)} & Rubric item &
50 & 3.3610 & 1.0400 & 0.9080\\

2025 & $C=1.0$, $U=0.9$, $P=\mathrm{true}$ &
GPT-4o & GPT-4-Turbo &
\shortstack{Problem\\(text)} & Rubric item &
21 & 0.3702 & 1.0649 & 0.9588\\

2026 & All &
\multicolumn{2}{c}{GPT-6 Astra} &
\shortstack{Whole exam\\(PDF)} & -- &
100.0 & 1.8533 & 0.9973 & 0.9375\\

2026 & Parts retained after review flags &
\multicolumn{2}{c}{GPT-6 Astra} &
\shortstack{Whole exam\\(PDF)} & Problem part &
62.8 & 1.1837 & 1.0114 & 0.9800\\
\end{tabular}
\end{ruledtabular}
\endgroup
\end{table*}

The 2024 selection retained 84.1\% of exams and raised $R^2$ only from
0.459 to 0.491. The 2025 policies produced markedly higher $R^2$ for
selected subtotals: 0.908 at a reported 50\% judgment coverage and
0.959 at 21\%. Those results include the positive-only acceptance
policies and binary scoring described above. The three 100\% points
do not represent repeated estimates of one configuration; they show
that rubric refinement did not improve the raw total-score regressions
in that study~\cite{Kortemeyer2025}.

In 2026, excluding review-recommended parts leaves 13453 of 21420
historical judgments, or 62.8\%. The resulting fit has offset 1.184
points, slope 1.011, and $R^2=0.980$. It contains 250 exams because
two have no retained part. This is not a count of 250 completely
automated examinations: retained observations may include only
a subset of the original work.

The problem-level results are listed in Table~\ref{tab:problems2026}.
Raw $R^2$ ranges from 0.858 to 0.930. The accepted-judgment coverage
is lowest for Problem~2, at 45.7\%, and ranges from 63.7\% to 71.7\%
for the other problems. Filtered $N$ also differs by scope. For example,
249 exams contribute to the filtered Problem~4 fit, while 250
contribute to the filtered grand total: an exam may retain work in
another problem even when no part of Problem~4 remains.

\begin{table*}
\caption{2026 comparisons by complete problem, before and after exclusion
of review-recommended parts. Coverage counts historical rubric judgments,
including zero-score work in the eligible denominator. A filtered exam
is included whenever at least one part remains within the stated problem.}
\label{tab:problems2026}
\begin{ruledtabular}
\begin{tabular}{llrrrrrr}
Problem & Scenario & $N$ & Offset & Slope & $R^2$ & Accepted judgments & Coverage (\%)\\
\hline
1 & All      & 252 & 0.8290 & 0.9860 & 0.9300 & 5796/5796 & 100.0\\
1 & Filtered & 250 & 0.4909 & 0.9844 & 0.9571 & 3945/5796 & 68.1\\
2 & All      & 252 & 0.2587 & 1.0064 & 0.8576 & 5040/5040 & 100.0\\
2 & Filtered & 233 & 0.2042 & 1.0409 & 0.9725 & 2304/5040 & 45.7\\
3 & All      & 252 & 0.9271 & 0.9516 & 0.9041 & 4788/4788 & 100.0\\
3 & Filtered & 250 & 0.3495 & 0.9926 & 0.9706 & 3048/4788 & 63.7\\
4 & All      & 252 & 0.5630 & 0.9540 & 0.9158 & 5796/5796 & 100.0\\
4 & Filtered & 249 & 0.3181 & 0.9984 & 0.9566 & 4156/5796 & 71.7\\
\end{tabular}
\end{ruledtabular}
\end{table*}

\section{Discussion}

The clearest result is the change in the raw, matched comparisons.
With the same student records retained, the 2026 workflow shows much
stronger association with TA scores for all three selected parts.
For the analytical and graphical parts, the slopes also move
substantially closer to one. The result therefore cannot be explained
solely by referring more cumbersome scoring decisions to humans. It is present
before review filtering.

The graphical task is especially informative. A $T$--$s$ diagram is not
just a collection of labeled marks. Its meaning depends on which
states are connected, the direction and shape of the paths, and the
relationship to the relevant isobars. In the earlier workflows, this
information had to survive an intermediate prose description before
the scoring model could use it. Once a branch or state relationship
was omitted or misdescribed, repeated scoring of the same transcript
could not recover it. The direct PDF workflow removes that particular
external transcription step and supplies the student's diagram
together with the reference and surrounding work. This is a plausible
contributor to the improvement, but the study does not isolate it
from changes in the underlying model or rubric.

The scope and representation of the available context are also
relevant. Both earlier scoring workflows supplied transcribed work
from all parts of a problem. In 2025, the rubric and requested
judgments were restricted to one part, while the surrounding
transcribed work remained available. The 2026 workflow supplied
the complete student and reference PDFs together and requested
all criterion awards in a single response. This made the original
handwriting and diagrams available alongside the full scoring
guidance, without a separate transcription stage. Such access can
help interpret work continued across pages and calculations that
depend on earlier results, although the present comparison does
not isolate its contribution from changes in the model or rubric.

The different confidence mechanisms also reveal different notions of
reliability. Repeated-score agreement tests whether the system is
consistent with itself. Psychometric agreement tests whether a
judgment fits a pattern across students and items. A review flag
identifies an explicit difficulty in reading or interpreting the
available evidence. Each can be useful, but none is a calibrated
probability that the award is correct. In particular, an interpretation
error shared across runs can evade a dispersion filter, and an unusual
but valid student solution need not fit the class-wide response pattern.

The filtered grand totals in 2025 and 2026 include different
amounts of accepted work for different students. This can
make the TA and AI subtotals vary together: examinations
with more retained work offer both graders more opportunities
to award points. The correlation can therefore partly
reflect how much work was included. Under the 2025
positive-only policy, every retained criterion receives
full AI credit. In a hypothetical case where the TAs award
credit for exactly half of the retained one-point criteria
in every examination, the AI and TA subtotals would be
perfectly proportional, giving $R^2=1$, although half of
the individual awards would disagree. The slope would
reveal the systematic difference between the totals.

Agreement between totals can also conceal disagreements
that cancel when points are added. For example, in a part
with four one-point criteria, the TAs might credit the
first two criteria and the AI the last two. Both would
award two points, despite disagreeing on every item.
Such cancellation is possible when entire parts are
retained, including both credited and uncredited items,
as in the 2026 analysis. We therefore interpret offset,
slope, and $R^2$ as descriptions of agreement between
the accepted subtotals, together with coverage and the
filtering policy. They do not by themselves establish
agreement on individual rubric judgments.

Counting historical rubric judgments gives coverage an operational
interpretation that a point-weighted measure would not have. It
estimates the fraction of specified scoring decisions that a review
policy leaves with the AI. However, examining a small diagram feature
may take longer than checking a numerical substitution, and reading
the surrounding work can serve several criteria at once. The 62.8\%
coverage should therefore not be reported as 62.8\% of staff time
saved. Measuring that outcome would require a separate study of the
human review process.

A recent study of AI scoring
of written physics explanations explicitly compared model performance
with agreement between independent human graders~\cite{ChenWan2025}.
For the present study, some residual differences may reflect different
interpretations of incomplete reasoning or different degrees of leniency
in awarding partial credit. Establishing how much of the disagreement has
this origin would require independent human rescoring and adjudication
of disputed cases. The present comparison therefore does not, by itself,
establish a ceiling on attainable AI--human agreement.

For physics assessment, the result supports continued use of rich
written tasks rather than restricting the examination to easily
machine-checkable responses. The available systems have become more
capable of evaluating the varied evidence that such tasks produce.
That development is useful only if the scoring criteria remain
educationally meaningful and the resulting awards can be inspected
and corrected by humans. Strong aggregate agreement does not remove
the need to resolve consequential errors in an individual student's
work.

\section{Limitations}

This is a longitudinal case study of one thermodynamics exam and a
consenting subset of its students. The three selected parts differ
in point value, length, and physical content as well as representation.
They illustrate analytical, graphical, and numerical demands, but do
not establish average performance for those categories across physics.
The common-cohort analysis also conditions on records available in
all three years; 100\% coverage within that cohort is not a measure
of end-to-end success over every original scan or examinee.

The workflows changed together with their models. Transcription,
rubric wording, partial-credit conventions, available context,
repeated scoring, and review policies all differed. The 2025 paper
also documents errors in the rubric for Part~1(b), while the current
prompt explicitly handles reference inconsistencies. Historical scores are retained rather than retrospectively
repaired. Model retirements limit the possibility of
rerunning the original scoring workflows, but archived
outputs can still be reanalyzed. These are
appropriate comparisons of the workflows that were used, but they
cannot attribute the observed changes to model capability alone.
Nor do the publication years establish the best model available
throughout each calendar year.

The total-score overview is less tightly controlled than the
selected-part comparison: the historical coefficients were not
completely recalculated against the reconstructed TA reference.
The available records do not fully document the 2025 acceptance
denominator or the refinement stage of every recovered average.
Reproduction also requires the documented gas-constant weighting
convention and the distinction between raw fractional scores and
thresholded selected scores.

TA scoring provides a relevant operational reference, but no
independent adjudication was conducted for the present comparison.
Agreement with that reference does not establish the correctness of
every award. The reported regressions do not measure the frequency
of large individual errors, fairness across student groups, or the
effectiveness of a human appeal process. 

The new full-cohort analysis retains one completed scoring response
per exam and does not report uncertainty intervals. We therefore
make no claim of statistical significance or a guaranteed error
bound for accepted scores.

Finally, while data sent to Azure API endpoints are contractually excluded from future model training,  a public exam and worked solution could potentially become part of the
material encountered by later models. Training-data exposure cannot
be established from the available records. Familiarity with the
problem would not by itself identify what a particular student
wrote, but its possible contribution cannot be separated here.
Fresh examinations and prospective evaluation would test how well
the observed progress generalizes.

\section{Conclusion}

Across successive studies of the same handwritten thermodynamics
exam, the practical performance of AI scoring changed substantially.
On fixed student cohorts, the unfiltered 2026 scores were much more
strongly associated with TA awards for all three selected tasks.
The largest improvement occurred for the graphical task, and the
slopes for the analytical and graphical tasks moved substantially
closer to one. These changes were visible before any work was
excluded by confidence filtering. They document progress in complete
scoring systems, including their models, input representations,
examination context, and rubrics.

Review recommendations provide a way to direct human attention to
selected parts of the work. Among the retained responses, some
disagreements may concern familiar decisions about incomplete reasoning,
partial credit, and consistent follow-through. Further improvement
can therefore involve both more capable scoring systems and clearer
agreement about how the rubric should be applied.

For physics educators, the practical opportunity is to use this
growing capability to support the assessment of derivations,
diagrams, and linked calculations. Instructors determine which
steps must be shown, how alternative methods earn credit, and how
an earlier error should affect later awards. Making these decisions
explicit, and revisiting them when disputed responses expose an
ambiguity, helps align scoring with the intended learning goals.
Two years have changed what a practical scoring system can do with
handwritten physics; deciding what deserves credit remains an
educational responsibility.

\section*{Data and code availability}
The data supporting the findings of this article cannot be made publicly available due to restrictions imposed by the human-subject research protocol. The code for this project is available from \url{https://gitlab.com/ethz-ai-for-education/ethel}.

\begin{acknowledgments}
Studies involving human subjects are reviewed by ETH Zurich's Ethics Committee; this study was approved as Study~2023-N-286. 
We would also like to thank Julian N{\"ohl} and his team for making the exam data available, as well as the students who participated in this study, and Daria Horica and Alina Yaroshchuk who assisted with logistics and the data analysis that laid the foundations for this study. This study is part of project Ethel~\cite{kortemeyer2024ethel}. While being the subject of this study, AI (GPT-6 Astra) was also used for a broad array of the tasks involved, including ideation, creation of analysis programs and graphs, literature review, writing, and revision/verification tasks.
\end{acknowledgments}

\bibliographystyle{apsrev4-2}
\bibliography{TwoYears}

%apsrev4-2.bst 2019-01-14 (MD) hand-edited version of apsrev4-1.bst
%Control: key (0)
%Control: author (72) initials jnrlst
%Control: editor formatted (1) identically to author
%Control: production of article title (-1) disabled
%Control: page (0) single
%Control: year (1) truncated
%Control: production of eprint (0) enabled
\begin{thebibliography}{30}%
\makeatletter
\providecommand \@ifxundefined [1]{%
 \@ifx{#1\undefined}
}%
\providecommand \@ifnum [1]{%
 \ifnum #1\expandafter \@firstoftwo
 \else \expandafter \@secondoftwo
 \fi
}%
\providecommand \@ifx [1]{%
 \ifx #1\expandafter \@firstoftwo
 \else \expandafter \@secondoftwo
 \fi
}%
\providecommand \natexlab [1]{#1}%
\providecommand \enquote  [1]{``#1''}%
\providecommand \bibnamefont  [1]{#1}%
\providecommand \bibfnamefont [1]{#1}%
\providecommand \citenamefont [1]{#1}%
\providecommand \href@noop [0]{\@secondoftwo}%
\providecommand \href [0]{\begingroup \@sanitize@url \@href}%
\providecommand \@href[1]{\@@startlink{#1}\@@href}%
\providecommand \@@href[1]{\endgroup#1\@@endlink}%
\providecommand \@sanitize@url [0]{\catcode `\\12\catcode `\$12\catcode
  `\&12\catcode `\#12\catcode `\^12\catcode `\_12\catcode `\%12\relax}%
\providecommand \@@startlink[1]{}%
\providecommand \@@endlink[0]{}%
\providecommand \url  [0]{\begingroup\@sanitize@url \@url }%
\providecommand \@url [1]{\endgroup\@href {#1}{\urlprefix }}%
\providecommand \urlprefix  [0]{URL }%
\providecommand \Eprint [0]{\href }%
\providecommand \doibase [0]{https://doi.org/}%
\providecommand \selectlanguage [0]{\@gobble}%
\providecommand \bibinfo  [0]{\@secondoftwo}%
\providecommand \bibfield  [0]{\@secondoftwo}%
\providecommand \translation [1]{[#1]}%
\providecommand \BibitemOpen [0]{}%
\providecommand \bibitemStop [0]{}%
\providecommand \bibitemNoStop [0]{.\EOS\space}%
\providecommand \EOS [0]{\spacefactor3000\relax}%
\providecommand \BibitemShut  [1]{\csname bibitem#1\endcsname}%
\let\auto@bib@innerbib\@empty
%</preamble>
\bibitem [{\citenamefont {Lin}(1982)}]{lin1982learning}%
  \BibitemOpen
  \bibfield  {author} {\bibinfo {author} {\bibfnamefont {H.}~\bibnamefont
  {Lin}},\ }\href@noop {} {\bibfield  {journal} {\bibinfo  {journal} {Physics
  teacher}\ }\textbf {\bibinfo {volume} {20}},\ \bibinfo {pages} {151}
  (\bibinfo {year} {1982})}\BibitemShut {NoStop}%
\bibitem [{\citenamefont {Burkholder}\ \emph {et~al.}(2020)\citenamefont
  {Burkholder}, \citenamefont {Miles}, \citenamefont {Layden}, \citenamefont
  {Wang}, \citenamefont {Fritz},\ and\ \citenamefont
  {Wieman}}]{Burkholder2020}%
  \BibitemOpen
  \bibfield  {author} {\bibinfo {author} {\bibfnamefont {E.~W.}\ \bibnamefont
  {Burkholder}}, \bibinfo {author} {\bibfnamefont {J.~K.}\ \bibnamefont
  {Miles}}, \bibinfo {author} {\bibfnamefont {T.~J.}\ \bibnamefont {Layden}},
  \bibinfo {author} {\bibfnamefont {K.~D.}\ \bibnamefont {Wang}}, \bibinfo
  {author} {\bibfnamefont {A.~V.}\ \bibnamefont {Fritz}},\ and\ \bibinfo
  {author} {\bibfnamefont {C.~E.}\ \bibnamefont {Wieman}},\ }\href
  {https://doi.org/10.1103/PhysRevPhysEducRes.16.010123} {\bibfield  {journal}
  {\bibinfo  {journal} {Phys. Rev. Phys. Educ. Res.}\ }\textbf {\bibinfo
  {volume} {16}},\ \bibinfo {pages} {010123} (\bibinfo {year}
  {2020})}\BibitemShut {NoStop}%
\bibitem [{\citenamefont {Docktor}\ \emph {et~al.}(2016)\citenamefont
  {Docktor}, \citenamefont {Dornfeld}, \citenamefont {Frodermann},
  \citenamefont {Heller}, \citenamefont {Hsu}, \citenamefont {Jackson},
  \citenamefont {Mason}, \citenamefont {Ryan},\ and\ \citenamefont
  {Yang}}]{Docktor2016}%
  \BibitemOpen
  \bibfield  {author} {\bibinfo {author} {\bibfnamefont {J.~L.}\ \bibnamefont
  {Docktor}}, \bibinfo {author} {\bibfnamefont {J.}~\bibnamefont {Dornfeld}},
  \bibinfo {author} {\bibfnamefont {E.}~\bibnamefont {Frodermann}}, \bibinfo
  {author} {\bibfnamefont {K.}~\bibnamefont {Heller}}, \bibinfo {author}
  {\bibfnamefont {L.}~\bibnamefont {Hsu}}, \bibinfo {author} {\bibfnamefont
  {K.~A.}\ \bibnamefont {Jackson}}, \bibinfo {author} {\bibfnamefont
  {A.}~\bibnamefont {Mason}}, \bibinfo {author} {\bibfnamefont {Q.~X.}\
  \bibnamefont {Ryan}},\ and\ \bibinfo {author} {\bibfnamefont
  {J.}~\bibnamefont {Yang}},\ }\href
  {https://doi.org/10.1103/PhysRevPhysEducRes.12.010130} {\bibfield  {journal}
  {\bibinfo  {journal} {Phys. Rev. Phys. Educ. Res.}\ }\textbf {\bibinfo
  {volume} {12}},\ \bibinfo {pages} {010130} (\bibinfo {year}
  {2016})}\BibitemShut {NoStop}%
\bibitem [{\citenamefont {Hull}\ \emph {et~al.}(2013)\citenamefont {Hull},
  \citenamefont {Kuo}, \citenamefont {Gupta},\ and\ \citenamefont
  {Elby}}]{Hull2013}%
  \BibitemOpen
  \bibfield  {author} {\bibinfo {author} {\bibfnamefont {M.~M.}\ \bibnamefont
  {Hull}}, \bibinfo {author} {\bibfnamefont {E.}~\bibnamefont {Kuo}}, \bibinfo
  {author} {\bibfnamefont {A.}~\bibnamefont {Gupta}},\ and\ \bibinfo {author}
  {\bibfnamefont {A.}~\bibnamefont {Elby}},\ }\href
  {https://doi.org/10.1103/PhysRevSTPER.9.010105} {\bibfield  {journal}
  {\bibinfo  {journal} {Phys. Rev. ST Phys. Educ. Res.}\ }\textbf {\bibinfo
  {volume} {9}},\ \bibinfo {pages} {010105} (\bibinfo {year}
  {2013})}\BibitemShut {NoStop}%
\bibitem [{\citenamefont
  {Kortemeyer}(2023{\natexlab{a}})}]{Kortemeyer2023Grading}%
  \BibitemOpen
  \bibfield  {author} {\bibinfo {author} {\bibfnamefont {G.}~\bibnamefont
  {Kortemeyer}},\ }\href {https://doi.org/10.1103/PhysRevPhysEducRes.19.020163}
  {\bibfield  {journal} {\bibinfo  {journal} {Phys. Rev. Phys. Educ. Res.}\
  }\textbf {\bibinfo {volume} {19}},\ \bibinfo {pages} {020163} (\bibinfo
  {year} {2023}{\natexlab{a}})}\BibitemShut {NoStop}%
\bibitem [{\citenamefont {Kortemeyer}\ \emph {et~al.}(2024)\citenamefont
  {Kortemeyer}, \citenamefont {N{\"o}hl},\ and\ \citenamefont
  {Onishchuk}}]{Kortemeyer2024}%
  \BibitemOpen
  \bibfield  {author} {\bibinfo {author} {\bibfnamefont {G.}~\bibnamefont
  {Kortemeyer}}, \bibinfo {author} {\bibfnamefont {J.}~\bibnamefont
  {N{\"o}hl}},\ and\ \bibinfo {author} {\bibfnamefont {D.}~\bibnamefont
  {Onishchuk}},\ }\href {https://doi.org/10.1103/PhysRevPhysEducRes.20.020144}
  {\bibfield  {journal} {\bibinfo  {journal} {Phys. Rev. Phys. Educ. Res.}\
  }\textbf {\bibinfo {volume} {20}},\ \bibinfo {pages} {020144} (\bibinfo
  {year} {2024})}\BibitemShut {NoStop}%
\bibitem [{\citenamefont {Kortemeyer}\ and\ \citenamefont
  {N{\"o}hl}(2025)}]{Kortemeyer2025}%
  \BibitemOpen
  \bibfield  {author} {\bibinfo {author} {\bibfnamefont {G.}~\bibnamefont
  {Kortemeyer}}\ and\ \bibinfo {author} {\bibfnamefont {J.}~\bibnamefont
  {N{\"o}hl}},\ }\href {https://doi.org/10.1103/PhysRevPhysEducRes.21.010136}
  {\bibfield  {journal} {\bibinfo  {journal} {Phys. Rev. Phys. Educ. Res.}\
  }\textbf {\bibinfo {volume} {21}},\ \bibinfo {pages} {010136} (\bibinfo
  {year} {2025})}\BibitemShut {NoStop}%
\bibitem [{\citenamefont {Chen}\ and\ \citenamefont {Wan}(2025)}]{ChenWan2025}%
  \BibitemOpen
  \bibfield  {author} {\bibinfo {author} {\bibfnamefont {Z.}~\bibnamefont
  {Chen}}\ and\ \bibinfo {author} {\bibfnamefont {T.}~\bibnamefont {Wan}},\
  }\href {https://doi.org/10.1103/PhysRevPhysEducRes.21.010126} {\bibfield
  {journal} {\bibinfo  {journal} {Phys. Rev. Phys. Educ. Res.}\ }\textbf
  {\bibinfo {volume} {21}},\ \bibinfo {pages} {010126} (\bibinfo {year}
  {2025})}\BibitemShut {NoStop}%
\bibitem [{\citenamefont {McGinness}\ and\ \citenamefont
  {Baumgartner}(2025)}]{mcginness2025can}%
  \BibitemOpen
  \bibfield  {author} {\bibinfo {author} {\bibfnamefont {L.}~\bibnamefont
  {McGinness}}\ and\ \bibinfo {author} {\bibfnamefont {P.}~\bibnamefont
  {Baumgartner}},\ }\href@noop {} {\bibfield  {journal} {\bibinfo  {journal}
  {Physical Review Physics Education Research}\ }\textbf {\bibinfo {volume}
  {21}},\ \bibinfo {pages} {020155} (\bibinfo {year} {2025})}\BibitemShut
  {NoStop}%
\bibitem [{\citenamefont {Mok}\ \emph {et~al.}(2025)\citenamefont {Mok},
  \citenamefont {Akhtar}, \citenamefont {Clare}, \citenamefont {Li},
  \citenamefont {Ida}, \citenamefont {Ross},\ and\ \citenamefont
  {Campanelli}}]{mok2025using}%
  \BibitemOpen
  \bibfield  {author} {\bibinfo {author} {\bibfnamefont {R.}~\bibnamefont
  {Mok}}, \bibinfo {author} {\bibfnamefont {F.}~\bibnamefont {Akhtar}},
  \bibinfo {author} {\bibfnamefont {L.}~\bibnamefont {Clare}}, \bibinfo
  {author} {\bibfnamefont {C.}~\bibnamefont {Li}}, \bibinfo {author}
  {\bibfnamefont {J.}~\bibnamefont {Ida}}, \bibinfo {author} {\bibfnamefont
  {L.}~\bibnamefont {Ross}},\ and\ \bibinfo {author} {\bibfnamefont
  {M.}~\bibnamefont {Campanelli}},\ }\href@noop {} {\bibfield  {journal}
  {\bibinfo  {journal} {Physics Education}\ }\textbf {\bibinfo {volume} {60}},\
  \bibinfo {pages} {035006} (\bibinfo {year} {2025})}\BibitemShut {NoStop}%
\bibitem [{\citenamefont {Henderson}\ \emph {et~al.}(2004)\citenamefont
  {Henderson}, \citenamefont {Yerushalmi}, \citenamefont {Kuo}, \citenamefont
  {Heller},\ and\ \citenamefont {Heller}}]{Henderson2004Grading}%
  \BibitemOpen
  \bibfield  {author} {\bibinfo {author} {\bibfnamefont {C.}~\bibnamefont
  {Henderson}}, \bibinfo {author} {\bibfnamefont {E.}~\bibnamefont
  {Yerushalmi}}, \bibinfo {author} {\bibfnamefont {V.~H.}\ \bibnamefont {Kuo}},
  \bibinfo {author} {\bibfnamefont {P.}~\bibnamefont {Heller}},\ and\ \bibinfo
  {author} {\bibfnamefont {K.}~\bibnamefont {Heller}},\ }\href
  {https://doi.org/10.1119/1.1634963} {\bibfield  {journal} {\bibinfo
  {journal} {American Journal of Physics}\ }\textbf {\bibinfo {volume} {72}},\
  \bibinfo {pages} {164} (\bibinfo {year} {2004})}\BibitemShut {NoStop}%
\bibitem [{\citenamefont {Slepkov}\ and\ \citenamefont
  {Shiell}(2014)}]{SlepkovShiell2014}%
  \BibitemOpen
  \bibfield  {author} {\bibinfo {author} {\bibfnamefont {A.~D.}\ \bibnamefont
  {Slepkov}}\ and\ \bibinfo {author} {\bibfnamefont {R.~C.}\ \bibnamefont
  {Shiell}},\ }\href {https://doi.org/10.1103/PhysRevSTPER.10.020120}
  {\bibfield  {journal} {\bibinfo  {journal} {Phys. Rev. ST Phys. Educ. Res.}\
  }\textbf {\bibinfo {volume} {10}},\ \bibinfo {pages} {020120} (\bibinfo
  {year} {2014})}\BibitemShut {NoStop}%
\bibitem [{\citenamefont {Speirs}\ \emph {et~al.}(2026)\citenamefont {Speirs},
  \citenamefont {Lane},\ and\ \citenamefont {Laird}}]{speirs2026thematic}%
  \BibitemOpen
  \bibfield  {author} {\bibinfo {author} {\bibfnamefont {J.~C.}\ \bibnamefont
  {Speirs}}, \bibinfo {author} {\bibfnamefont {W.~B.}\ \bibnamefont {Lane}},\
  and\ \bibinfo {author} {\bibfnamefont {N.}~\bibnamefont {Laird}},\
  }\href@noop {} {\bibfield  {journal} {\bibinfo  {journal} {Physical Review
  Physics Education Research}\ }\textbf {\bibinfo {volume} {22}},\ \bibinfo
  {pages} {010115} (\bibinfo {year} {2026})}\BibitemShut {NoStop}%
\bibitem [{\citenamefont {Chasteen}\ \emph {et~al.}(2012)\citenamefont
  {Chasteen}, \citenamefont {Pepper}, \citenamefont {Caballero}, \citenamefont
  {Pollock},\ and\ \citenamefont {Perkins}}]{ChasteenEtAl2012}%
  \BibitemOpen
  \bibfield  {author} {\bibinfo {author} {\bibfnamefont {S.~V.}\ \bibnamefont
  {Chasteen}}, \bibinfo {author} {\bibfnamefont {R.~E.}\ \bibnamefont
  {Pepper}}, \bibinfo {author} {\bibfnamefont {M.~D.}\ \bibnamefont
  {Caballero}}, \bibinfo {author} {\bibfnamefont {S.~J.}\ \bibnamefont
  {Pollock}},\ and\ \bibinfo {author} {\bibfnamefont {K.~K.}\ \bibnamefont
  {Perkins}},\ }\href {https://doi.org/10.1103/PhysRevSTPER.8.020108}
  {\bibfield  {journal} {\bibinfo  {journal} {Phys. Rev. ST Phys. Educ. Res.}\
  }\textbf {\bibinfo {volume} {8}},\ \bibinfo {pages} {020108} (\bibinfo {year}
  {2012})}\BibitemShut {NoStop}%
\bibitem [{\citenamefont {Paul}\ \emph {et~al.}(2014)\citenamefont {Paul},
  \citenamefont {Potter},\ and\ \citenamefont
  {Weiss}}]{Paul2014ResponseCategory}%
  \BibitemOpen
  \bibfield  {author} {\bibinfo {author} {\bibfnamefont {C.}~\bibnamefont
  {Paul}}, \bibinfo {author} {\bibfnamefont {W.}~\bibnamefont {Potter}},\ and\
  \bibinfo {author} {\bibfnamefont {B.}~\bibnamefont {Weiss}},\ }\href
  {https://doi.org/10.1119/1.4897587} {\bibfield  {journal} {\bibinfo
  {journal} {The Physics Teacher}\ }\textbf {\bibinfo {volume} {52}},\ \bibinfo
  {pages} {485} (\bibinfo {year} {2014})}\BibitemShut {NoStop}%
\bibitem [{\citenamefont
  {Kortemeyer}(2023{\natexlab{b}})}]{kortemeyer2023could}%
  \BibitemOpen
  \bibfield  {author} {\bibinfo {author} {\bibfnamefont {G.}~\bibnamefont
  {Kortemeyer}},\ }\href@noop {} {\bibfield  {journal} {\bibinfo  {journal}
  {Physical Review Physics Education Research}\ }\textbf {\bibinfo {volume}
  {19}},\ \bibinfo {pages} {010132} (\bibinfo {year}
  {2023}{\natexlab{b}})}\BibitemShut {NoStop}%
\bibitem [{\citenamefont {Levine}\ \emph {et~al.}(2027)\citenamefont {Levine},
  \citenamefont {Aenlle}, \citenamefont {Zilles}, \citenamefont {West},\ and\
  \citenamefont {Silva}}]{levine2026automated}%
  \BibitemOpen
  \bibfield  {author} {\bibinfo {author} {\bibfnamefont {J.}~\bibnamefont
  {Levine}}, \bibinfo {author} {\bibfnamefont {M.}~\bibnamefont {Aenlle}},
  \bibinfo {author} {\bibfnamefont {C.}~\bibnamefont {Zilles}}, \bibinfo
  {author} {\bibfnamefont {M.}~\bibnamefont {West}},\ and\ \bibinfo {author}
  {\bibfnamefont {M.}~\bibnamefont {Silva}},\ }in\ \href@noop {} {\emph
  {\bibinfo {booktitle} {Artificial Intelligence in Education}}},\ Vol.\
  \bibinfo {volume} {16582}\ (\bibinfo {organization} {Springer},\ \bibinfo
  {year} {2027})\ pp.\ \bibinfo {pages} {627--636}\BibitemShut {NoStop}%
\bibitem [{\citenamefont {Polverini}\ and\ \citenamefont
  {Gregorcic}(2024{\natexlab{a}})}]{polverini2024evaluating}%
  \BibitemOpen
  \bibfield  {author} {\bibinfo {author} {\bibfnamefont {G.}~\bibnamefont
  {Polverini}}\ and\ \bibinfo {author} {\bibfnamefont {B.}~\bibnamefont
  {Gregorcic}},\ }in\ \href@noop {} {\emph {\bibinfo {booktitle} {Frontiers in
  Education}}},\ Vol.~\bibinfo {volume} {9}\ (\bibinfo {organization}
  {Frontiers Media SA},\ \bibinfo {year} {2024})\ p.\ \bibinfo {pages}
  {1452414}\BibitemShut {NoStop}%
\bibitem [{\citenamefont {Polverini}\ and\ \citenamefont
  {Gregorcic}(2024{\natexlab{b}})}]{polverini2024performance}%
  \BibitemOpen
  \bibfield  {author} {\bibinfo {author} {\bibfnamefont {G.}~\bibnamefont
  {Polverini}}\ and\ \bibinfo {author} {\bibfnamefont {B.}~\bibnamefont
  {Gregorcic}},\ }\href@noop {} {\bibfield  {journal} {\bibinfo  {journal}
  {Physical review physics education research}\ }\textbf {\bibinfo {volume}
  {20}},\ \bibinfo {pages} {010109} (\bibinfo {year}
  {2024}{\natexlab{b}})}\BibitemShut {NoStop}%
\bibitem [{\citenamefont {Kortemeyer}\ \emph {et~al.}(2025)\citenamefont
  {Kortemeyer}, \citenamefont {Babayeva}, \citenamefont {Polverini},
  \citenamefont {Widenhorn},\ and\ \citenamefont
  {Gregorcic}}]{kortemeyer2025multilingual}%
  \BibitemOpen
  \bibfield  {author} {\bibinfo {author} {\bibfnamefont {G.}~\bibnamefont
  {Kortemeyer}}, \bibinfo {author} {\bibfnamefont {M.}~\bibnamefont
  {Babayeva}}, \bibinfo {author} {\bibfnamefont {G.}~\bibnamefont {Polverini}},
  \bibinfo {author} {\bibfnamefont {R.}~\bibnamefont {Widenhorn}},\ and\
  \bibinfo {author} {\bibfnamefont {B.}~\bibnamefont {Gregorcic}},\ }\href@noop
  {} {\bibfield  {journal} {\bibinfo  {journal} {Physical Review Physics
  Education Research}\ }\textbf {\bibinfo {volume} {21}},\ \bibinfo {pages}
  {020101} (\bibinfo {year} {2025})}\BibitemShut {NoStop}%
\bibitem [{\citenamefont {Polverini}\ and\ \citenamefont
  {Gregorcic}(2025)}]{polverini2025multimodal}%
  \BibitemOpen
  \bibfield  {author} {\bibinfo {author} {\bibfnamefont {G.}~\bibnamefont
  {Polverini}}\ and\ \bibinfo {author} {\bibfnamefont {B.}~\bibnamefont
  {Gregorcic}},\ }\href@noop {} {\bibfield  {journal} {\bibinfo  {journal}
  {European Journal of Physics}\ }\textbf {\bibinfo {volume} {46}},\ \bibinfo
  {pages} {055708} (\bibinfo {year} {2025})}\BibitemShut {NoStop}%
\bibitem [{\citenamefont {{OpenAI}}(2024)}]{OpenAI2024GPT4o}%
  \BibitemOpen
  \bibfield  {author} {\bibinfo {author} {\bibnamefont {{OpenAI}}},\ }\href
  {https://openai.com/index/hello-gpt-4o/} {\bibinfo {title} {Hello {GPT-4o}}}
  (\bibinfo {year} {2024}),\ \bibinfo {note} {published May 13,
  2024}\BibitemShut {NoStop}%
\bibitem [{\citenamefont {{OpenAI}}(2023{\natexlab{a}})}]{OpenAI2023GPT4}%
  \BibitemOpen
  \bibfield  {author} {\bibinfo {author} {\bibnamefont {{OpenAI}}},\ }\href
  {https://doi.org/10.48550/arXiv.2303.08774} {\bibinfo {title} {{GPT-4}
  technical report}} (\bibinfo {year} {2023}{\natexlab{a}}),\ \Eprint
  {https://arxiv.org/abs/2303.08774} {arXiv:2303.08774 [cs.CL]} \BibitemShut
  {NoStop}%
\bibitem [{\citenamefont {{OpenAI}}(2023{\natexlab{b}})}]{OpenAI2023Vision}%
  \BibitemOpen
  \bibfield  {author} {\bibinfo {author} {\bibnamefont {{OpenAI}}},\ }\href
  {https://openai.com/index/gpt-4v-system-card/} {\bibinfo {title}
  {{GPT-4V(ision)} system card}} (\bibinfo {year} {2023}{\natexlab{b}}),\
  \bibinfo {note} {published September 25, 2023}\BibitemShut {NoStop}%
\bibitem [{\citenamefont {Liu}\ \emph {et~al.}(2024)\citenamefont {Liu},
  \citenamefont {Lin}, \citenamefont {Hewitt}, \citenamefont {Paranjape},
  \citenamefont {Bevilacqua}, \citenamefont {Petroni},\ and\ \citenamefont
  {Liang}}]{Liu2024LostMiddle}%
  \BibitemOpen
  \bibfield  {author} {\bibinfo {author} {\bibfnamefont {N.~F.}\ \bibnamefont
  {Liu}}, \bibinfo {author} {\bibfnamefont {K.}~\bibnamefont {Lin}}, \bibinfo
  {author} {\bibfnamefont {J.}~\bibnamefont {Hewitt}}, \bibinfo {author}
  {\bibfnamefont {A.}~\bibnamefont {Paranjape}}, \bibinfo {author}
  {\bibfnamefont {M.}~\bibnamefont {Bevilacqua}}, \bibinfo {author}
  {\bibfnamefont {F.}~\bibnamefont {Petroni}},\ and\ \bibinfo {author}
  {\bibfnamefont {P.}~\bibnamefont {Liang}},\ }\href
  {https://doi.org/10.1162/tacl_a_00638} {\bibfield  {journal} {\bibinfo
  {journal} {Transactions of the Association for Computational Linguistics}\
  }\textbf {\bibinfo {volume} {12}},\ \bibinfo {pages} {157} (\bibinfo {year}
  {2024})}\BibitemShut {NoStop}%
\bibitem [{\citenamefont {Hsieh}\ \emph {et~al.}(2024)\citenamefont {Hsieh},
  \citenamefont {Sun}, \citenamefont {Kriman}, \citenamefont {Acharya},
  \citenamefont {Rekesh}, \citenamefont {Jia}, \citenamefont {Zhang},\ and\
  \citenamefont {Ginsburg}}]{Hsieh2024RULER}%
  \BibitemOpen
  \bibfield  {author} {\bibinfo {author} {\bibfnamefont {C.-P.}\ \bibnamefont
  {Hsieh}}, \bibinfo {author} {\bibfnamefont {S.}~\bibnamefont {Sun}}, \bibinfo
  {author} {\bibfnamefont {S.}~\bibnamefont {Kriman}}, \bibinfo {author}
  {\bibfnamefont {S.}~\bibnamefont {Acharya}}, \bibinfo {author} {\bibfnamefont
  {D.}~\bibnamefont {Rekesh}}, \bibinfo {author} {\bibfnamefont
  {F.}~\bibnamefont {Jia}}, \bibinfo {author} {\bibfnamefont {Y.}~\bibnamefont
  {Zhang}},\ and\ \bibinfo {author} {\bibfnamefont {B.}~\bibnamefont
  {Ginsburg}},\ }in\ \href {https://openreview.net/forum?id=kIoBbc76Sy} {\emph
  {\bibinfo {booktitle} {First Conference on Language Modeling}}}\ (\bibinfo
  {year} {2024})\ \Eprint {https://arxiv.org/abs/2404.06654} {arXiv:2404.06654
  [cs.CL]} \BibitemShut {NoStop}%
\bibitem [{\citenamefont {Geifman}\ and\ \citenamefont
  {El-Yaniv}(2017)}]{Geifman2017}%
  \BibitemOpen
  \bibfield  {author} {\bibinfo {author} {\bibfnamefont {Y.}~\bibnamefont
  {Geifman}}\ and\ \bibinfo {author} {\bibfnamefont {R.}~\bibnamefont
  {El-Yaniv}},\ }in\ \href
  {https://proceedings.neurips.cc/paper/2017/hash/4a8423d5e91fda00bb7e46540e2b0cf1-Abstract.html}
  {\emph {\bibinfo {booktitle} {Advances in Neural Information Processing
  Systems}}},\ Vol.~\bibinfo {volume} {30}\ (\bibinfo {year}
  {2017})\BibitemShut {NoStop}%
\bibitem [{Note1()}]{Note1}%
  \BibitemOpen
  \bibinfo {note} {For Part~3(a), we assign the combined 2025 gas-constant
  criterion a weight of 1.5 points: one point for the expression and half a
  point for its numerical evaluation in the original examination
  rubric.}\BibitemShut {Stop}%
\bibitem [{Note2()}]{Note2}%
  \BibitemOpen
  \bibinfo {note} {The 2026 TA totals are calculated from the detailed grading
  sheets. Seven detailed-sheet totals in the full source workbook differ from
  its summary totals. The historical total-score reference vectors have not all
  been reconstructed to check these discrepancies.}\BibitemShut {Stop}%
\bibitem [{\citenamefont {Kortemeyer}(2024)}]{kortemeyer2024ethel}%
  \BibitemOpen
  \bibfield  {author} {\bibinfo {author} {\bibfnamefont {G.}~\bibnamefont
  {Kortemeyer}},\ }\href@noop {} {\bibfield  {journal} {\bibinfo  {journal}
  {The Physics Teacher}\ }\textbf {\bibinfo {volume} {62}},\ \bibinfo {pages}
  {698} (\bibinfo {year} {2024})}\BibitemShut {NoStop}%
\end{thebibliography}%

\end{document}